\documentclass[11pt,a4paper]{article}
\pdfoutput=1

\usepackage{jheppub}
\usepackage{amsmath,amsfonts,amssymb,amsthm}
\usepackage{url}

\numberwithin{equation}{section}

\newcommand{\fft}[2]{\frac{#1}{#2}}
\newcommand{\ft}[2]{{\textstyle\frac{#1}{#2}}}
\newcommand{\nn}{\nonumber}

\DeclareMathOperator{\Tr}{Tr}

\preprint{LCTP-24-11}

\title{Higher derivative heterotic supergravity on a torus and supersymmetry}

\author[a]{Sabarenath Jayaprakash}

\emailAdd{sabare@umich.edu}

\author[a]{and James T. Liu}

\emailAdd{jimliu@umich.edu}

\affiliation[a]{Leinweber Center for Theoretical Physics, University of Michigan, Ann Arbor, MI 48109, U.S.A.}

\abstract{Ignoring ten-dimensional heterotic gauge fields, heterotic supergravity reduced on a $d$-dimensional torus gives rise to a half-maximal supergravity coupled to $d$ vector multiplets.  The reduced theory has a continuous $O(d,d;\mathbb R)/O(d)_-\times O(d)_+$ symmetry that persists to all perturbative orders in the string $\alpha'$ expansion.  We highlight this symmetry by explicitly reducing the bosonic sector of four-derivative heterotic supergravity as well as its fermionic supersymmetry variations.  After appropriate field redefinitions, the resulting action and supersymmetry variations are manifestly $O(d)_-\times O(d)_+$ invariant.  This reduction allows us to explore the interplay between the gravity and vector multiplets beyond leading order, where (in our conventions) $O(d)_-$ is the supergravity R-symmetry while $O(d)_+$ is a flavor symmetry of the $d$ vector multiplets.}


\arxivnumber{}

\begin{document}

\maketitle


\section{Introduction}

One of the features of string theory that distinguishes it from field theories of point particles is the presence of a string length scale $\ell_s=\sqrt{\alpha'}$.  While low energy observers may not have direct access to the string scale, one may nevertheless view string theory as an effective supergravity coupled to matter theory.  In this sense, the extended nature of stringy behavior manifests itself through the $\alpha'$ expansion, and probes of string theory turn into studies of the effective field theory with higher derivative couplings.

From the bottom up, one may in principle write down very general effective actions.  But it is not at all obvious whether such generic actions have a UV completion in string theory.  In fact, one may expect the symmetries of string theory to impose stringent conditions on the Wilson coefficients of the higher dimensional operators in the effective action.  More generally, we may ask the question of whether there are general features of low energy actions arising from string theory that can yield insights on their stringy origins.

One of the striking symmetries of compactified string theory is T-duality, which can be understood as a gauge symmetry of string theory.  As an exact symmetry of string theory, this imposes strong constraints on the form of the effective field theory action and can be used to restrict the form of the higher derivative couplings \cite{Godazgar:2013bja,Marques:2015vua,Hohm:2015doa,Garousi:2019wgz,Garousi:2019mca,Garousi:2020gio,Codina:2020kvj,David:2021jqn,David:2022jcl}.  Perhaps the simplest framework for exploring T-duality is a torus compactification.  For the heterotic string on $T^d$, which we focus on here, the T-duality group is the discrete $O(d,d+16;\mathbb Z)$ group.  However, in the low energy effective action, this is enhanced to a continuous $O(d,d+16;\mathbb R)$ global symmetry of the massless bosonic degrees of freedom.  In particular, the scalars parametrize an $O(d,d+16;\mathbb R)/O(d)\times O(d+16)$ coset with corresponding $O(d,d+16;\mathbb Z)$ T-duality group, while the fermions transform locally under $O(d)\times O(d+16)$.

While it is known that the $O(d,d+16;\mathbb R)$ symmetry of the heterotic action persists to all orders in the $\alpha'$ expansion, the actual structure of the higher-derivative couplings are not always apparent.  A fruitful approach to constructing T-duality invariant actions is through double field theory \cite{Siegel:1993th,Hull:2009mi,Hohm:2010jy,Hohm:2010pp} and its higher derivative extensions
\cite{Hohm:2013jaa,Bedoya:2014pma,Hohm:2014xsa,Coimbra:2014qaa,Lee:2015kba,Marques:2015vua,Baron:2017dvb,Lescano:2021guc}.  However, even in this case, the structure of the dimensionally reduced actions can be rather involved.  Moreover, one generally needs to resort to field redefinitions in order to make the transformation properties manifest.  At $\mathcal O(\alpha')$, explicit $O(d,d;\mathbb R)$ invariant dimensionally reduced bosonic and heterotic string effective actions were constructed in \cite{Eloy:2020dko,Elgood:2020xwu,Ortin:2020xdm}.   This construction involved a large set of non-trivial field redefinitions to put the action into a form that does not involve any additional derivatives on the fields beyond that expected from gauge covariance.

In this paper, we take a complementary approach by highlighting the local $O(d)\times O(d)$ invariance of the truncated heterotic theory reduced on a torus.  The motivation here is to highlight the supersymmetry structure of the lower-dimensional supergravity and to make a direct connection to the reduced supersymmetry variations.  Heterotic supergravity without gauge fields reduces on $T^d$ to half-maximal supergravity in $10-d$ dimensions coupled to $d$ vector multiplets where one of the $O(d)\times O(d)$ factors is identified as the R-symmetry and the other as a flavor symmetry.  Correspondingly, the reduced fermions split into a gravitino and dilatino in the gravity multiplet and $d$ gaugini in the vector multiplets.

The dimensional reduction makes use of the torsionful connections, $\Omega_\pm=\omega\pm\fft12\mathcal H$, where $\omega$ is the conventional spin connection and $\mathcal H^{\alpha\beta}\equiv H_\mu{}^{\alpha\beta}dx^\mu$.  This torsion connection arises naturally in the four-derivative heterotic action constructed in \cite{Bergshoeff:1988nn,Bergshoeff:1989de}, and is furthermore incorporated in higher derivative approaches to double field theory \cite{Marques:2015vua,Baron:2017dvb} and torus reductions \cite{Elgood:2020xwu,Ortin:2020xdm,David:2021jqn,David:2022jcl}.  Using $\Omega_\pm$, we find that many terms in the reduction manifestly transform covariantly under $O(d)\times O(d)$.  At the same time, there are a set of terms that do not, but a relatively straightforward $\mathcal O(\alpha')$ field redefinition can be used to eliminate them.

In addition to the $O(d)\times O(d)$ invariant four-derivative bosonic action, we highlight the supersymmetry variation of the fermions.  Here additional work is involved in obtaining suitable field redefinitions of the fermions so that they split cleanly between the gravity and vector multiplets.  In principle, one would have to reduce the $\alpha'$ corrected fermionic action to confirm that these redefinitions lead to manifest $O(d)\times O(d)$ invariance of the fermion sector.  However, they are consistent with the results of \cite{Liu:2023fqq}, where the vector multiplets were truncated out.

The outline of this paper is as follows.  In the following section, we present our conventions with an overview of heterotic supergravity and the leading order torus reduction.  In section~\ref{sec:alpha'}, we turn to the reduction of the $\mathcal O(\alpha')$ Lagrangian as well as the supersymmetry variations of the fermions.  In section~\ref{sec:redefine} we perform additional field redefinitions to remove explicit derivative terms from the reduced Lagrangian to facilitate comparison with the $O(d,d;\mathbb R)$ invariant form of the Lagrangian of \cite{Eloy:2020dko}.  Finally, we present some concluding remarks in section~\ref{sec:discuss}.


\section{Torus reduction of heterotic supergravity}

The heterotic string is a ten-dimensional half-maximally supersymmetric string theory.  In its low energy limit its massless degrees of freedom are well described by heterotic supergravity with the gravity fields $(g_{MN}, B_{MN},\phi,\psi_M,\lambda)$ coupled to $E_8\times E_8$ or $SO(32)$ matter, $(A_M^a,\chi^a)$.  We will truncate out the heterotic gauge fields and restrict ourselves to the sector.  In the string frame, the bosonic Lagrangian, including four-derivative corrections, can be written as \cite{Bergshoeff:1988nn,Bergshoeff:1989de,Metsaev:1987zx,Chemissany:2007he}
\begin{equation}
    e^{-1}\mathcal L_{\mathrm{het}}=e^{-2\phi}\left(R+4(\partial_M\phi)^2-\fft1{12}(\tilde H_{MNP})^2+\fft{\alpha'}8(R_{MN}{}^{AB}(\Omega_+))^2\right).
\label{eq:hetlag}
\end{equation}
The structure of this Lagrangian is quite remarkable.  Here $R_{MN}{}^{AB}(\Omega_+)$ is the torsionful Riemann tensor
\begin{equation}
    R(\Omega_+)=d\Omega_++\Omega_+\wedge\Omega_+,
\end{equation}
built from the torsionful connection
\begin{equation}
    \Omega_\pm=\omega\pm\fft12\tilde{\mathcal H},\qquad\tilde{\mathcal H}^{AB}\equiv\tilde H_M{}^{AB}dx^M.
\end{equation}
In the above, $\tilde H_{MNP}$ is the three-form with Chern-Simons addition
\begin{equation}
    \tilde H=H-\fft{\alpha'}4\omega_{3L}(\Omega_+),\qquad H=dB,
\label{eq:tildeH}
\end{equation}
where
\begin{equation}
    \omega_{3L}(\Omega_+)=\Tr\left(\Omega_+\wedge d\Omega_++\fft23\Omega_+\wedge\Omega_+\wedge\Omega_+\right).
\end{equation}
In particular, $\tilde H$ satisfies the Bianchi identity
\begin{equation}
    d\tilde H=-\fft{\alpha'}4\Tr\left(R(\Omega_+)\wedge R(\Omega_+)\right).
\end{equation}

Note that $\tilde H$ appears on both sides of (\ref{eq:tildeH}) since it enters implicitly on the right-hand side through the definition of the torsionful connection $\Omega_+$.  This structure can be seen from the supersymmetric construction of \cite{Bergshoeff:1988nn,Bergshoeff:1989de} as well as from $\sigma$-model anomaly considerations \cite{Sen:1985tq}.  The proper way to make sense of this expression for $\tilde H$ is as an iterative expansion in $\alpha'$.  Since we work only to first order in $\alpha'$, it is sufficient for our purposes to take the leading order expression for the torsionful connection, so that
\begin{equation}
    \tilde H=H-\fft{\alpha'}4\omega_{3L}(\omega+\ft12\mathcal H),\qquad\mathcal H=H_M{}^{AB}dx^M.
\end{equation}
Similar considerations apply when working with the torsionful Riemann tensor.

While (\ref{eq:hetlag}) is a four-derivative Lagrangian, the $\mathcal O(\alpha')$ equations of motion take the simple form \cite{Bergshoeff:1988nn,Bergshoeff:1989de}
\begin{align}
    0&=R-4(\partial_M\phi)^2+4\Box\phi-\fft1{12}(\tilde H_{MNP})^2+\fft{\alpha'}8(R_{MNAB}(\Omega_+))^2,\nn\\
    0&=R_{MN}+2\nabla_M\nabla_N\phi-\fft14\tilde H_{MAB}\tilde H_N{}^{AB}+\fft{\alpha'}4R_{MPAB}(\Omega_+)R_N{}^{PAB}(\Omega_+),\nn\\
    0&=\nabla^M(e^{-2\phi}\tilde H_{MNP}),
\end{align}
where all covariant derivatives are with respect to the ten-dimensional metric, $g_{MN}$.

The form of the bosonic action ensures that it is actually supersymmetric to $\mathcal O(\alpha'^2)$ \cite{Bergshoeff:1988nn,Bergshoeff:1989de}, where the fermion supersymmetry variations are built from the connection $\Omega_-$
\begin{align}
    \delta_\epsilon\psi_M&=\nabla_M(\Omega_-)\epsilon=\left(\partial_M+\fft14\Omega_{-\,M}{}^{AB}\Gamma_{AB}\right)\epsilon=\left(\nabla_M-\fft18\tilde H_{MNP}\Gamma^{NP}\right)\epsilon,\nn\\
    \delta_\epsilon\lambda&=\left(\Gamma^M\partial_M\phi-\fft1{12}\tilde H_{MNP}\Gamma^{MNP}\right)\epsilon.
\label{eq:susies}
\end{align}
The interplay between $\Omega_+$ in the Lagrangian and $\Omega_-$ in the variations ensures invariance under supersymmetry.

\subsection{The reduction ansatz}

We start with a conventional torus reduction of the heterotic Lagrangian by taking the Kaluza-Klein ansatz
\begin{align}
    ds^2&=g_{\mu\nu}dx^\mu dx^\nu+g_{ij}\eta^i\eta^j,\qquad\eta^i=dy^i+A_\mu^idx^\mu,\nn\\
    B&=\fft12b_{\mu\nu}dx^\mu\wedge dx^\nu+B_{\mu i}dx^\mu\wedge\eta^i+\fft12b_{ij}\eta^i\wedge\eta^j,\nn\\
    \phi&=\varphi+\fft14\log\det g_{ij}.
\label{eq:redans}
\end{align}
It follows that the three-form field strength can be written as
\begin{equation}
    H=dB=\fft16h_{\mu\nu\rho}dx^\mu\wedge dx^\nu\wedge dx^\rho+\tilde G_i\wedge\eta^i+\fft12db_{ij}\eta^i\wedge\eta^j,
\end{equation}
where
\begin{equation}
    h=db-B_i\wedge F^i,\qquad\tilde G_i=dB_i-b_{ij}F^j.
\end{equation}
Truncating to the zero-mode sector on the torus is guaranteed to yield a consistent reduction.  Thus one may simply insert the above reduction ansatz into the heterotic Lagrangian, (\ref{eq:hetlag}).  At the two-derivative level, the reduced Lagrangian take the form \cite{Maharana:1992my}
\begin{align}
    e^{-1}\mathcal L_{\partial^2}&=e^{-2\varphi}\biggl(R+4(\partial_\mu\varphi)^2-\fft1{12}(h_{\mu\nu\lambda})^2-\fft14\left(g_{ij}F_{\mu\nu}^iF^{\mu\nu\,j}+g^{ij}\tilde G_{\mu\nu\,i}\tilde G^{\mu\nu}_j\right)\nn\\
    &\kern4em-\fft14\Tr\left(g^{-1}\partial_\mu gg^{-1}\partial^\mu g-g^{-1}\partial_\mu bg^{-1}\partial^\mu b\right)\biggr).
\end{align}

To demonstrate the $O(d,d;\mathbb R)$ invariance of this Lagrangian, we introduce the $O(d,d)$ matrix
\begin{equation}
    \mathcal H=\begin{pmatrix}g-bg^{-1}b&bg^{-1}\\-g^{-1}b&g^{-1}\end{pmatrix},
\label{eq:mathcalH}
\end{equation}
where $g$ and $b$ are the internal components, $g_{ij}$ and $b_{ij}$, respectively.  Note that $\mathcal H^{-1}=\eta\mathcal H\eta$, where
\begin{equation}
    \eta=\begin{pmatrix}0&1\\1&0\end{pmatrix},
\end{equation}
is the $O(d,d)$ metric.  The reduced Lagrangian can now be written in terms of $\mathcal H$ as
\begin{align}
    e^{-1}\mathcal L_{\partial^2}&=e^{-2\varphi}\left(R+4(\partial_\mu\varphi)^2-\fft1{12}(h_{\mu\nu\lambda})^2-\fft14\mathcal F_{\mu\nu}\mathcal H\mathcal F^{\mu\nu}+\fft18\Tr(\partial_\mu\mathcal H\eta\partial^\mu\mathcal H\eta)\right),
\label{eq:Oddlag2}
\end{align}
where we have grouped the vector fields as an $O(d,d)$ vector
\begin{equation}
    \mathcal F=\begin{pmatrix}F^i\\G_i\end{pmatrix}=\begin{pmatrix}dA^i\\dB_i\end{pmatrix}.
\end{equation}
In this form, the Lagrangian, (\ref{eq:Oddlag2}), is manifestly invariant under $O(d,d;\mathbb R)$ transformations of the form
\begin{equation}
    \mathcal H\to\Omega^{-1\,T}\mathcal H\Omega^{-1},\qquad\mathcal F\to\Omega\mathcal F,
\end{equation}
where $\Omega$ is an $O(d,d;\mathbb R)$ matrix satisfying $\Omega^T\eta\Omega=\eta$ \cite{Maharana:1992my}.

\subsection{Generalized vielbein basis}

As is well known, the reduced scalars transform under the coset $O(d,d)/O(d)_-\times O(d)_+$, the gauge fields transform under $O(d,d)$, and the fermions transform under the bottom of the coset.  To highlight this structure, we introduce a generalized vielbein basis.  The starting point is the natural vielbein basis corresponding to the torus reduced metric, (\ref{eq:redans})
\begin{equation}
    E_M{}^A=\begin{pmatrix}e^\alpha{}_\mu&A_\mu^ie^a{}_i\\0&e^a{}_i\end{pmatrix}.
\end{equation}
The internal vielbein components, $e^a{}_i$, can be combined with the internal $B$-field components $b_{ij}$ to form the $O(d,d)$ vielbein
\begin{equation}
    V^A{}_I=\begin{pmatrix}v^{(-)\,a}{}_i&v^{(-)\,ai}\\v^{(+)\,a}{}_i&v^{(+)\,ai}\end{pmatrix}=\begin{pmatrix}e^a{}_i+e^{aj}b_{ji}\quad&-e^{ai}\\e^a{}_i-e^{aj}b_{ji}&e^{ai}\end{pmatrix},
\end{equation}

\begin{equation}
    (V^{-1})^I_A=\begin{pmatrix}v^{(-)\,i}{}_a&v^{(+)\,i}{}_a\\v^{(-)}_{ia}&v^{(+)}_{ia}\end{pmatrix}=\fft12\begin{pmatrix}e^i{}_a&e^i{}_a\\-e_{ia}+b_{ij}e^j{}_a\quad&e_{ia}+b_{ij}e^j{}_a\end{pmatrix}.
\end{equation}
Note that we have chosen a normalization convention so as to avoid $\sqrt2$ factors in the vielbein.

The $O(d,d)$ vielbein and its inverse can be written in a matrix notation
\begin{equation}
    V=\begin{pmatrix}e+e^{-1T}b&-e^{-1T}\\e-e^{-1T}b\quad&e^{-1T}\end{pmatrix},\qquad V^{-1}=\fft12\begin{pmatrix}e^{-1}&e^{-1}\\-e^T+be^{-1}\quad&e^T+be^{-1}\end{pmatrix},
\label{eq:genviel}
\end{equation}
and it satisfies the relation
\begin{equation}
    \mathcal H=\fft12V^TV,
\end{equation}
where $\mathcal H$ is the scalar matrix defined in (\ref{eq:mathcalH}).  Given the generalized vielbein, we can introduce the $O(d)_-\times O(d)_+$ components of the vector fields, $\mathcal F^A=V\mathcal F$, or
\begin{equation}
    \begin{pmatrix}F^{(-)}\\F^{(+)}\end{pmatrix}=V\begin{pmatrix}F^i\\G_i\end{pmatrix}=\begin{pmatrix}e^a{}_iF^i-e^{ai}\tilde G_i\\e^a{}_iF^i+e^{ai}\tilde G_i\end{pmatrix}.
\end{equation}

Given the vielbein, we can calculate the Mauer-Cartan form
\begin{equation}
    L=dVV^{-1}=Q+P=\begin{pmatrix}Q^{(--)}&0\\0&Q^{(++)}\end{pmatrix}+\begin{pmatrix}0&P^{(-+)}\\P^{(+-)}&0\end{pmatrix},
\end{equation}
where
\begin{align}
    Q^{(--)}&=\fft12e^{-1T}(e^Tde-de^Te+db)e^{-1},\nn\\
    Q^{(++)}&=\fft12e^{-1T}(e^Tde-de^Te-db)e^{-1},\nn\\
    P^{(-+)\,T}=P^{(+-)}&=\fft12e^{-1T}d(g-b)e^{-1}.
\end{align}
As usual, $Q^{(--)}$ and $Q^{(++)}$ form a composite $O(d)_-\times O(d)_+$ connection, while $P^{(\pm\mp)}$ is used for the scalar kinetic term in the Lagrangian.  From $dL-L\wedge L=0$, we obtain the identities
\begin{align}
    &dQ^{(\mp\mp)}-Q^{(\mp\mp)}\wedge Q^{(\mp\mp)}=P^{(\mp\pm)}\wedge P^{(\pm\mp)},\nn\\
    &dP^{(\pm\mp)}-P^{(\pm\mp)}\wedge Q^{(\mp\mp)}-Q^{(\pm\pm)}\wedge P^{(\pm\mp)}=0.
\end{align}
The second line takes the form $\mathcal DP=0$ where $\mathcal D$ is the covariant derivative with respect to the composite connection $Q$.  These identities will be useful below when simplifying the four-derivative Lagrangian.

\subsection{The two-derivative reduction and supersymmetry}

Using the generalized vielbein, the $O(d,d)$ form of the two-derivative Lagrangian, (\ref{eq:Oddlag2}), can be rewritten in terms of $O(d)_-\times O(d)_+$ fields as
\begin{align}
    e^{-1}\mathcal L_{\partial^2}
    &=e^{-2\varphi}\biggl(R+4(\partial_\mu\varphi)^2-\fft1{12}h_{\mu\nu\rho}^2-\fft18(F_{\mu\nu}^{(-)\,a})^2\nn\\
    &\kern10em-\fft18(F_{\mu\nu}^{(+)\,a})^2-P_\mu^{(+-)\,ab}P^{(+-)\,\mu\,ab}\biggr).
\label{eq:Lagd2}
\end{align}
Here we have split the Lagrangian into two parts; the first line corresponds to the lower-dimensional gravity multiplet, while the second line corresponds to the lower-dimensional vector multiplets.

The identification of the gravity and vector multiplets is more apparent when we examine the reduced leading order fermion supersymmetry transformations
\begin{align}
    \delta_\epsilon\psi_\mu&=\left[\nabla_\mu(\omega_-)-\fft14Q_\mu^{(--)\,ab}\Gamma^{ab}+\fft14F_{\mu\nu}^{(-)\,a}\Gamma^\nu\Gamma^a\right]\epsilon,\nn\\
    \delta_\epsilon\psi_i&=e_i^a\left[-\fft12P_\mu^{(+-)\,ab}\Gamma^\mu\Gamma^b-\fft18F_{\mu\nu}^{(+)\,a}\Gamma^{\mu\nu}\right]\epsilon,\nn\\
    \delta\tilde\lambda&=\left[\Gamma^\mu\partial_\mu\varphi-\fft1{12}h_{\mu\nu\rho}\Gamma^{\mu\nu\rho}+\fft18F_{\mu\nu}^{(-)\,a}\Gamma^{\mu\nu}\Gamma^a\right]\epsilon.
\label{eq:losusy}
\end{align}
Here $\tilde\lambda$ is the shifted dilatino, $\tilde\lambda=\lambda-\Gamma^i\psi_i$.  This is a direct counterpart of the dilaton shift in the torus reduction, (\ref{eq:redans}).  Note that the spinors transform under $O(d)_-$ while the gaugini transform as a vector of $O(d)_+$.

Before proceeding to the $\mathcal O(\alpha')$ reduction, we present the reduced equations of motion%
\footnote{From now on, we will mostly use lower greek indices and upper roman indices to streamline the notation.  Repeated indices are to be contracted as usual with implicit use of the appropriate metric.}
\begin{align}
    \mathcal{E}_{g,\alpha\beta}\;\, =&\; R(\omega)_{\alpha\beta} +2 \nabla_\alpha\nabla_\beta \varphi - \dfrac{1}{4} \left(F^{(+)\, a}_{\alpha\gamma} F^{(+)\, a}_{\beta\gamma} + F^{(-)\, a}_{\alpha\gamma} F^{(-)\, a}_{\beta\gamma} \right) -\dfrac{1}{4} h_{\alpha\gamma\delta} \, h_{\beta\gamma\delta}\nn\\
    &\;  - P_\alpha^{(+-)\, ab} P_\beta^{(+-)\, ab}\nn\\
    \mathcal{E}_{F^+,\alpha a} = & \; e^{2\varphi} \,\nabla_\beta(e^{-2\varphi} F_{\alpha\beta}^{(+)\, a}) - Q_\beta^{(++)\, ab} F_{\alpha\beta}^{(+)\, b} - \dfrac{1}{2} h_{\alpha\beta\delta}F_{\beta\delta}^{(+)\,a} + P_\beta^{(+-)\,ac} F_{\alpha\beta}^{(-)\, c}\nn\\
    \mathcal{E}_{F^-,\alpha a} = &\; e^{2\varphi} \,\nabla_\beta(e^{-2\varphi} F_{\alpha\beta}^{(-)\, a}) - Q_\beta^{(--)\, ab} F_{\alpha\beta}^{(-)\, b} + \dfrac{1}{2} h_{\alpha\beta\delta}F_{\beta\delta}^{(-)\,a} + P_\beta^{(+-)\,ca} F_{\alpha\beta}^{(+)\, c}\nn\\
    \mathcal{E}_{P,ab}\ \ \, = & \; e^{2\varphi} \nabla_\alpha(e^{-2\varphi} P_\alpha^{(+-)\, ab}) - \dfrac{1}{4} F^{(+)\, a}_{\alpha\beta} F^{(-)\, b}_{\alpha\beta}\nn\\
    \mathcal{E}_{h, \alpha\beta}\;\; \, =&\; e^{2\varphi} \nabla_\gamma (e^{-2\varphi}h_{\alpha\beta\gamma})\nn\\
    \mathcal{E}_{\phi}\ \ \ \  = &\; R(\omega) + 4 \Box \varphi - 4 (\nabla_\alpha \varphi)^2  - \dfrac{1}{8} \left(F^{(+)\, a}_{\alpha\beta} F^{(+)\, a}_{\alpha\beta} + F^{(-)\, a}_{\alpha\beta} F^{(-)\, a}_{\alpha\beta} \right) - \dfrac{1}{12} h_{\alpha\beta\gamma}^2\nn\\
    &\; - P_\alpha^{(+-)\, ab} P_\alpha^{(+-)\, ab}.
\label{eq:loeom}
\end{align}
These two-derivative equations of motion will be useful when performing field redefinitions of the $\mathcal O(\alpha')$ Lagrangian.

\section{The \texorpdfstring{$\mathcal O(\alpha')$}{O(alpha')} reduction}
\label{sec:alpha'}

We now turn to the $\mathcal O(\alpha')$ reduction of the heterotic Lagrangian, (\ref{eq:hetlag}).  Pulling out a factor of $\alpha'/8$, we write
\begin{equation}
    \mathcal L_{\mathrm{het}}=\mathcal L^{(0)}+\fft{\alpha'}8\mathcal L^{(1)},
\end{equation}
where
\begin{equation}
    e^{-1}\mathcal L^{(1)}=e^{-2\varphi}\left[(R_{MNPQ}(\Omega_+))^2+\fft13H_{MNP}\,\omega_{3L}(\Omega_+)_{MNP}\right].
\end{equation}
In principle, for a torus reduction, there is no need to correct the leading order Kaluza-Klein ansatz, (\ref{eq:redans}), since taking a full set of zero modes on the torus is sufficient to ensure consistency.  However, since we wish to highlight the $O(d,d)/O(d)_-\times O(d)_+$ structure of the reduced action, we find a field redefinition is needed to make this symmetry manifest.  As a result, the reduced four-derivative Lagrangian takes the form
\begin{equation}
    \mathcal L_{\partial^4}=\mathcal L^{(1)}+\delta\mathcal L^{(0)},
\end{equation}
where $\mathcal L^{(1)}$ is the reduction of the heterotic four-derivative couplings on the leading order ansatz, and $\delta\mathcal L^{(0)}$ is obtained from inserting the $\mathcal O(\alpha')$ corrected reduction ansatz into the heterotic two-derivative Lagrangian.

Since the four-derivative Lagrangian $\mathcal L^{(1)}$ involves the curvature of the torsionful connection, $\Omega_+$, as well as the Chern-Simons form, we first reduce $\Omega_+$ using the standard Kaluza-Klein ansatz, (\ref{eq:redans}).  We then compute various internal, external and mixed components of the Riemann tensor and Chern-Simons form.  For completeness, the expressions are written out in Appendix~\ref{app:Omega}.  As in \cite{Liu:2023fqq}, we shift the torsionful Chern-Simons form by a total derivative
\begin{equation}
    \tilde\omega_{3L}(\Omega_+)=\omega_{3L}(\Omega_+)+d\left(\fft12\omega_+^{\alpha\beta}F_{\alpha\beta}^{(-)\,c}\eta^c+F_{\mu\alpha}^{(+)\,b}P_\alpha^{(+-)\,bd}dx^\mu\wedge\eta^d\right),
\end{equation}
The first term in the total derivative removes a non-covariant term in the reduced Lagrangian and has the same form as that of \cite{Liu:2023fqq}.  While the second term is not strictly necessary, it removes some explicit derivative and non $O(d)_-\times O(d)_+$ invariant terms from the Lagrangian.  This shift of the Chern-Simons term corresponds to a field redefinition of $B_{\mu i}$ in the reduction ansatz, (\ref{eq:redans}).

Given the reduced torsionful Riemann tensor, (\ref{eq:Riemann+}), and the shifted Chern-Simons form, (\ref{eq:LCS+}), it is straightforward although tedious to write out the reduced four-derivative Lagrangian, $\mathcal L^{(1)}$.  We find that this Lagrangian will include both $O(d)_-\times O(d)_+$ invariant and non-invariant terms.  For example, the spacetime components of the torsionful Riemann tensor are
\begin{equation}
    R_{\gamma\delta}{}^{\alpha\beta}(\Omega_+)=R_{\gamma\delta}{}^{\alpha\beta}(\omega_+)-\fft12F_{[\gamma|\alpha}^{(+)\,c}F_{\delta]\beta}^{(+)\,c}-\fft14F_{\alpha\beta}^{(-)\,c}F_{\gamma\delta}^{(-)\,c}-\textcolor{red}{\fft14F_{\alpha\beta}^{(-)\,c}F_{\gamma\delta}^{(+)\,c}}.
\end{equation}
The first three terms are $O(d)_-\times O(d)_+$ singlets, but the last term is non-invariant.  Of course, the issue here is not that the reduced Lagrangian is non-invariant under T-duality, but rather that it lacks manifest $O(d_-)\times O(d)_+$ invariance when applying the leading order torus reduction, (\ref{eq:redans}).  As we demonstrate here, manifest $O(d)_-\times O(d)_+$ invariance can be recovered by a suitable field redefinition, corresponding to an $\mathcal O(\alpha')$ modification of the reduction ansatz.

Following \cite{Liu:2023fqq}, we conjecture that a shift of the internal metric, $g_{ij}\to g_{ij}+\delta g_{ij}$ is sufficient.  Since we are interested in the fermions transformations as well, we parametrize this shift as
\begin{equation}
    \delta e_i^a=T^{ab}e_i^b,\qquad\delta g_{ij}=2e_i^aT^{ab}e_j^b,
\label{eq:Tabshift}
\end{equation}
where $T^{ab}$ is a symmetric matrix in tangent space.  Inserting this shift into the leading order Lagrangian, (\ref{eq:Lagd2}), results in
\begin{align}
    e^{-1}\delta\mathcal L^{(0)}&=e^{-2\varphi}\biggl[\left(-\fft12F_{\mu\nu}^{(+)\,a}F_{\mu\nu}^{(-)\,b}+2P_\mu^{(+-)\,ca}P_\mu^{(+-)\,cb}-2P_\mu^{(+-)\,ac}P_\mu^{(+-)\,cb}\right)T^{(--)\,ab}\nn\\
    &\kern4em-P_\mu^{(+-)\,ab}\mathcal D_\mu T^{(--)\,ab}\biggr].
\end{align}
Note that here we have assumed that both indices on $T^{ab}$ transform under $O(d)_-$.  By explicit computation, we find that the non-invariant terms can be removed by taking
\begin{equation}
    T^{ab}=\fft1{32}F_{\mu\nu}^{(-)\,a}F_{\mu\nu}^{(-)\,b}+\fft14P_\mu^{(+-)\,ca}P_\mu^{(+-)\,cb}.
\label{eq:Tab}
\end{equation}
The resulting four-derivative Lagrangian then takes the form
\begin{equation}
    \mathcal L_{\partial^4}=\mathcal L_1+\mathcal L_2+\mathcal L_3,
\label{eq:lag4d123}
\end{equation}
where
\begin{align}
    e^{-1}\mathcal L_1&=e^{-2\varphi}\biggl[(R_{\alpha\beta\gamma\delta}(\omega_+))^2+\fft13h_{\alpha\beta\gamma}\omega_{3L}(\omega_+)_{\alpha\beta\gamma}-R_{\alpha\beta\gamma\delta}(\omega_+)F_{\alpha\beta}^{(-)\,a}F_{\gamma\delta}^{(-)\,a}\nn\\
    &\kern4em+\fft12(\mathcal D_\gamma^{(+)}F_{\alpha\beta}^{(-)\,a})^2+\fft18F_{\alpha\beta}^{(-)\,a}F_{\beta\gamma}^{(-)\,a}F_{\gamma\delta}^{(-)\,b}F_{\delta\alpha}^{(-)\,b}-\fft18F_{\alpha\beta}^{(-)\,a}F_{\beta\gamma}^{(-)\,b}F_{\gamma\delta}^{(-)\,a}F_{\delta\alpha}^{(-)\,b}\nn\\
    &\kern4em+\fft18F_{\alpha\beta}^{(-)\,a}F_{\alpha\beta}^{(-)\,b}F_{\gamma\delta}^{(-)\,a}F_{\gamma\delta}^{(-)\,b}\biggr],
\label{eq:L_1}
\end{align}
represents four-derivative couplings of the gravity multiplet,
\begin{align}
    e^{-1}\mathcal L_2&=e^{-2\varphi}\biggl[\fft18F_{\alpha\beta}^{(+)\,a}F_{\alpha\beta}^{(+)\,b}F_{\gamma\delta}^{(+)\,a}F_{\gamma\delta}^{(+)\,b}-\fft14F_{\alpha\beta}^{(+)\,a}F_{\beta\gamma}^{(+)\,b}F_{\gamma\delta}^{(+)\,a}F_{\delta\alpha}^{(+)\,b}\nn\\
    &\kern4em+\fft18F_{\alpha\beta}^{(+)\,a}F_{\beta\gamma}^{(+)\,a}F_{\gamma\delta}^{(+)\,b}F_{\delta\alpha}^{(+)\,b}+(\mathcal D_\mu^{\prime(+)}F_{\nu\textcolor{cyan}{\gamma}}^{(+)\,a})^2-\mathcal D_\mu^{\prime(+)}F_{\nu\textcolor{cyan}{\gamma}}^{(+)\,a}\mathcal D_\nu^{\prime(+)}F_{\mu\textcolor{cyan}{\gamma}}^{(+)\,a}\nn\\
    &\kern4em+P_\gamma^{(+-)\,ac}P_\gamma^{(+-)\,bc}F_{\alpha\beta}^{(+)\,a}F_{\alpha\beta}^{(+)\,b} -3 P_\alpha^{(+-)\,ac}P_\beta^{(+-)\,bc}F_{\alpha\gamma}^{(+)\,b}F_{\beta\gamma}^{(+)\,a}\nn\\
    &\kern4em+P_\alpha^{(+-)\,ac}P_\beta^{(+-)\,bc}F_{\alpha\gamma}^{(+)\,a}F_{\beta\gamma}^{(+)\,b}+P_\alpha^{(+-)\,ab}P_\beta^{(+-)\,ab}F_{\alpha\gamma}^{(+)\,c}F_{\beta\gamma}^{(+)\,c}\nn\\
    &\kern4em+2P_\alpha^{(+-)\,ab}P_\alpha^{(+-)\,cb}P_\beta^{(+-)\,cd}P_\beta^{(+-)\,ad} + 6P_\alpha^{(+-)\,ab}P_\beta^{(+-)\,cb}P_\beta^{(+-)\,cd}P_\alpha^{(+-)\,ad}\nn\\
    &\kern4em+2P_\alpha^{(+-)\,ab}P_\beta^{(+-)\,ab}P_\alpha^{(+-)\,cd}P_\beta^{(+-)\,cd}-6P_\alpha^{(+-)\,ab}P_\beta^{(+-)\,cb}P_\alpha^{(+-)\,cd}P_\beta^{(+-)\,ad}\nn\\
    &\kern4em+4(\mathcal D_\gamma^{(+)}P_\alpha^{(+-)\,bd})^2\biggr]
\label{eq:L_2}
\end{align}
represents four-derivative couplings of the vector multiplets, and
\begin{align}
    e^{-1}\mathcal L_3&= e^{-2\varphi}\biggl[\fft13h_{\alpha\mu\nu}(\omega_3(-Q^{(++)})_{\alpha\mu\nu}-3F_{\alpha\gamma}^{(+)\,a}\mathcal D_\mu^{\prime(+)}F_{\nu\textcolor{cyan}{\gamma}}^{(+)\,a})-R_{\alpha\beta\gamma\delta}(\omega_+)F_{\alpha\gamma}^{(+)\,a}F_{\beta\delta}^{(+)\,a}\nn\\
    &\kern4em+\fft12F_{\alpha\beta}^{(-)\,a}F_{\beta\gamma}^{(+)\,b}F_{\gamma\delta}^{(-)\,a}F_{\delta\alpha}^{(+)\,b}+\fft14F_{\alpha\beta}^{(-)\,a}F_{\beta\gamma}^{(-)\,a}F_{\gamma\delta}^{(+)\,b}F_{\delta\alpha}^{(+)\,b}\nn\\
    &\kern4em+\fft32P_\gamma^{(+-)\,ca}P_\gamma^{(+-)\,cb}F_{\alpha\beta}^{(-)\,a}F_{\alpha\beta}^{(-)\,b}+P_\alpha^{(+-)\,ab}P_\beta^{(+-)\,ab}F_{\alpha\gamma}^{(-)\,c}F_{\beta\gamma}^{(-)\,c}\nn\\
    &\kern4em+P_\alpha^{(+-)\,ca}P_\beta^{(+-)\,cb}F_{\alpha\gamma}^{(-)\,a}F_{\beta\gamma}^{(-)\,b}-2P_\alpha^{(+-)\,ca}P_\beta^{(+-)\,cb}F_{\alpha\gamma}^{(-)\,b}F_{\beta\gamma}^{(-)\,a}\nn\\
    &\kern4em-4P_\gamma^{(+-)\,ab}F_{\mu\nu}^{(-)\,b}\mathcal D_\mu^{\prime(+)}F_{\nu\textcolor{cyan}{\gamma}}^{(+)\,a}-2P_\beta^{(+-)\,ad}F_{\gamma\alpha}^{(+)\,a}\mathcal D_\gamma^{(+)} F_{\alpha\beta}^{(-)\,d}\nn\\
    &\kern4em-2F_{\alpha\beta}^{(-)\,d}F_{\gamma\beta}^{(+)\,b}\mathcal D_\gamma^{(+)}P_\alpha^{(+-)\,bd}\biggr]
\label{eq:L_3}
\end{align}
are mixed gravity-vector couplings.  Here the derivative $\mathcal D^{(+)}$ is covariant with respect to the torsionful connection $\omega_+=\omega+\fft12h$ and the composite connection $Q^{(++)}$, while $\mathcal D'^{(+)}$ acts according to
\begin{equation}
    \mathcal D_\mu'^{(+)}F_{\nu\textcolor{cyan}{\gamma}}^{(+)\,a}=\partial_\mu F_{\nu\textcolor{cyan}{\gamma}}^{(+)\,a}-\Gamma^\lambda{}_{\mu\nu}F_{\lambda\textcolor{cyan}{\gamma}}^{(+)\,a}+\omega_{+\,\mu\textcolor{cyan}{\gamma}}{}^{\textcolor{cyan}{\delta}} F_{\nu\textcolor{cyan}{\delta}}^{(+)\,a}-Q_\mu^{(++)\,ab}F_{\nu\textcolor{cyan}{\gamma}}^{(+)\,b}.
\end{equation}
The $O(d)_-\times O(d)_+$ invariance of this four-derivative Lagrangian is manifest as all derivatives are covariant under the $Q$ connection and $+$ and $-$ indices contract in pairs.  The four-derivative gravity multiplet couplings, (\ref{eq:L_1}), were previously obtained in \cite{Liu:2023fqq}.  Note that, at this point, the Lagrangian contains explicit derivatives of the field strengths.  These can be removed by further field redefinitions, and we will address this below.  However, we first consider the supersymmetry variations at $\mathcal O(\alpha')$.

\subsection{The \texorpdfstring{$\mathcal O(\alpha')$}{O(alpha')} supersymmetry variations}

While our treatment of the fermionic sector is incomplete, it is useful to work out the supersymmetry variations of the fermionic fields, as they are of use in constructing and studying BPS configurations of the reduced theory.  The starting point are the ten-dimensional gravitino and dilatino variations given in (\ref{eq:susies}).  Note that there the $\mathcal O(\alpha')$ corrections are entirely contained in the definition of $\tilde H_{MNP}$.  However, for the torus reduction it is useful to make these corrections explicit, so that
\begin{align}
    \delta_\epsilon\psi_M&=\delta_\epsilon\psi_M^{(0)}+\alpha' \delta_\epsilon\psi_M^{(1)},\nn\\
    \delta_\epsilon\lambda&=\delta_\epsilon\lambda^{(0)}+\alpha'\delta_\epsilon\lambda^{(1)}.
\label{eq:susyalpha'}
\end{align}
The ten-dimensional gravitino, $\psi_M$, reduces to the lower-dimensional gravitino, $\psi_\mu$, along with a set of gaugini, $\psi_i$, while the ten-dimensional dilatino, $\lambda$ reduces to the lower-dimensional dilatino, $\tilde\lambda$.  At leading order, the variations are given in (\ref{eq:losusy}), while at $\mathcal O(\alpha')$, the variations receive contributions from both the explicit terms in (\ref{eq:susyalpha'}) and the shift of the leading order variations arising from the $T^{ab}$ shift in (\ref{eq:Tabshift}).

We start with the lower-dimensional gravitino variation.  For the shift of the leading-order variation, we insert (\ref{eq:Tabshift}) into (\ref{eq:losusy}) to obtain
\begin{equation}
    \delta(\delta_\epsilon\psi_\mu^{(0)})=\left[-\fft12T^{ac}P_\mu^{(+-)\,cb}\Gamma^{ab}+\fft14F_{\mu\alpha}^{(+)\,c}T^{ac}\Gamma^\alpha\Gamma^a\right]\epsilon,
\end{equation}
where $T^{ab}$ is given in (\ref{eq:Tab}).  These terms are all non-invariant under $O(d)_+\times O(d)_-$, but cancel against similar non-invariant terms in $\delta_\epsilon\psi_\mu^{(1)}$.  The result is
\begin{align}
    \delta_\epsilon\psi_\mu^{(1)}+\delta(\delta_\epsilon\psi_\mu^{(0)})&=\fft1{32}\biggl[\left(\omega_{3L}(\omega_+)_{\mu\alpha\beta}+\omega_3(Q^{(++)})_{\mu\alpha\beta}-3F_{[\mu|\delta|}^{(+)\,a}\mathcal D_\alpha^{\prime(+)}F_{\beta]\delta}^{(+)\,a}\right)\Gamma^{\alpha\beta}\nn\\
    &\qquad+\Bigl(2R_{\mu\alpha\gamma\delta}(\omega_+)F_{\gamma\delta}^{(-)\,b}+8P_\delta^{(+-)\,cb}\mathcal D_{[\mu}^{\prime(+)}F_{\alpha]\delta}^{(+)\,c}-\fft14F_{\delta\epsilon}^{(-)\,b}F_{\delta\epsilon}^{(-)\,d}F_{\mu\alpha}^{(-)\,d}\nn\\
    &\kern4em-2P_\delta^{(+-)\,db}P_\delta^{(+-)\,de}F_{\mu\alpha}^{(-)\,e}+F_{\delta\epsilon}^{(-)\,b}F_{[\mu|\epsilon|}^{(+)\,d}F_{\alpha]\delta}^{(+)\,d}\Bigr)\Gamma^\alpha\Gamma^b\nn\\
    &\qquad+\Bigl(-4P_\delta^{(+-)\,db}\mathcal D_\mu^{(+)}P_\delta^{(+-)\,da}-\fft12F_{\delta\epsilon}^{(-)\,b}\mathcal D_\mu^{(+)}F_{\delta\epsilon}^{(-)\,a}\nn\\
    &\kern4em-2F_{\delta\epsilon}^{(-)\,a}F_{\mu\delta}^{(+)\,d}P_\epsilon^{(+-)\,db}\Bigr)\Gamma^{ab}\biggr]\epsilon.
\label{eq:deltagrav}
\end{align}
The full gravitino variation is then the combination of this expression along with $\delta_\epsilon\psi_\mu^{(0)}$ given in (\ref{eq:losusy}).  This is manifestly $O(d)_-\times O(d)_+$ covariant with all spinors transforming under $O(d)_-$.

Turning to the dilatino, we have the shift in the leading order variation
\begin{equation}
    \delta(\delta_\epsilon\tilde\lambda^{(0)})=\fft18F_{\alpha\beta}^{(+)\,b}T^{ab}\Gamma^{\alpha\beta}\Gamma^a\epsilon.
\end{equation}
This cancels the non-invariant terms in $\delta_\epsilon\tilde\lambda^{(1)}$, leaving
\begin{align}
    \delta_\epsilon\tilde\lambda^{(1)}+\delta(\delta_\epsilon\tilde\lambda^{(0)})&=\fft1{48}\biggl[\left(\omega_{3L}(\omega_+)_{\alpha\beta\gamma}+\omega_3(Q^{(++)})_{\alpha\beta\gamma}-3F_{\alpha\delta}^{(+)\,a}\mathcal D_\beta^{\prime(+)}F_{\gamma\delta}^{(+)\,a}\right)\Gamma^{\alpha\beta\gamma}\nn\\
    &\qquad+\Bigl(\fft32R_{\alpha\beta\gamma\delta}(\omega_+)F_{\gamma\delta}^{(-)\,c}+6P_\delta^{(+-)\,dc}\mathcal D_\alpha^{\prime(+)}F_{\beta\delta}^{(+)\,d}-\fft3{16}F_{\delta\epsilon}^{(-)\,c}F_{\delta\epsilon}^{(-)\,d}F_{\alpha\beta}^{(-)\,d}\nn\\
    &\kern4em-\fft32P_\delta^{(+-)\,dc}P_\delta^{(+-)\,de}F_{\alpha\beta}^{(-)\,e}+\fft34F_{\delta\epsilon}^{(-)\,c}F_{\alpha\epsilon}^{(+)\,d}F_{\beta\delta}^{(+)\,d}\Bigr)\Gamma^{\alpha\beta}\Gamma^c\nn\\
    &\qquad+\left(\fft14F_{\alpha\beta}^{(-)\,a}F_{\beta\gamma}^{(-)\,b}F_{\gamma\alpha}^{(-)\,c}-3F_{\delta\epsilon}^{(-)\,a}P_\epsilon^{(+-)\,db}P_\delta^{(+-)\,dc}\right)\Gamma^{abc}    \biggr]\epsilon.
\label{eq:deltadil}
\end{align}
Again, this is to be combined with $\delta_\epsilon\tilde\lambda^{(0)}$ in (\ref{eq:losusy}), and the full expression is $O(d)_-\times O(d)_+$ covariant.

We find that the gaugino variation, $\delta_\epsilon\psi_a$, is more involved, as here $a$ is naturally an $O(d)_+$ index, while the reduction of the ten-dimensional Lorentz-Chern-Simons form that appears in $\delta_\epsilon\psi_a^{(1)}$ preferentially transforms under $O(d)_-$.  In order to obtain an $O(d)_-\times O(d)_+$ covariant variation, we have to remove various $O(d)_-$ indices from the Lorentz-Chern-Simons term.  We start by defining a shifted gaugino variation $\delta_\epsilon\psi_a'$ according to
\begin{equation}
    \delta_\epsilon\psi_a'\equiv\delta(\delta_\epsilon\psi_a^{(0)})+\delta_\epsilon\psi_a^{(1)}-\fft18F_{\mu\nu}^{(-)\,a}[\hat{\mathcal D}_\mu,\hat{\mathcal D}_\nu]\epsilon-\fft12P_\mu^{(+-)\,ba}e^i_b[\hat{\mathcal D}_\mu,\hat{\mathcal D}_i]\epsilon.
\label{eq:psiaprime}
\end{equation}
Here we have defined the lowest order supercovariant derivatives
\begin{align}
    \hat{\mathcal D}_\mu&=\nabla_\mu(\omega_-)-\fft14Q^{(--)\,ab}\Gamma^{ab}+\fft14F_{\mu\nu}^{(-)\,a}\Gamma^\nu\Gamma^a,\nn\\
    \hat{\mathcal D}_i&=e_i^a\left(-\fft12P_\mu^{(+-)\,ab}\Gamma^\mu\Gamma^b-\fft18F_{\mu\nu}^{(+)\,a}\Gamma^{\mu\nu}\right).
\end{align}
Note that the $a$ index in (\ref{eq:psiaprime}) should be an $O(d)_+$ flavor index, so the two commutator terms on the right-hand side are designed to remove $O(d)_-$ terms from the Lorenz-Chern-Simons terms in $\delta_\epsilon\psi_a^{(1)}$.

We now present the explicit terms on the right-hand-side of (\ref{eq:psiaprime}).  For $\delta(\delta_\epsilon\psi_a^{(0)})$, we compute
\begin{equation}
    \delta(\delta_\epsilon\psi_a^{(0)})=\fft1{32}\left(-4T^{ab}F_{\alpha\beta}^{(-)\,b}\Gamma^{\alpha\beta}-16(\mathcal D_\alpha T^{ab}+(P_\alpha^{(+-)\,ca}-P_\alpha^{(+-)\,ac})T^{cb})\Gamma^\alpha\Gamma^b\right)\epsilon.
\end{equation}
For $\delta_\epsilon\psi_a^{(1)}$, we have
\begin{equation}
    \delta_\epsilon\psi_a^{(1)}=\fft1{32}\left(\tilde\omega_{3L}(\Omega_+)_{a\alpha\beta}\Gamma^{\alpha\beta}+2\tilde\omega_{3L}(\Omega_+)_{a\alpha b}\Gamma^\alpha\Gamma^b+\tilde\omega_{3L}(\Omega_+)_{abc}\Gamma^{bc}\right)\epsilon,
\end{equation}
where the components of the Lorentz-Chern-Simons form are given in (\ref{eq:LCS+}).  For the commutator of supercovariant derivatives, we have
\begin{align}
    [\hat{\mathcal D}_\mu,\hat{\mathcal D}_\nu]&=\left(\fft14R_{\mu\nu\alpha\beta}(\omega_-)-\fft18F_{\mu\alpha}^{(-)\,c}F_{\nu\beta}^{(-)\,c}\right)\Gamma^{\alpha\beta}+\fft12\mathcal D_{[\mu}^{\prime(-)}F_{\nu]\alpha}^{(-)\,b}\Gamma^\alpha\Gamma^b\nn\\
    &\qquad+\left(-\fft12P_{[\mu}^{(+-)\,db}P_{\nu]}^{(+-)\,dc}-\fft18F_{\mu\alpha}^{(-)\,b}F_{\nu\alpha}^{(-)\,c}\right)\Gamma^{bc},\nn\\
    [\hat{\mathcal D}_\mu,\hat{\mathcal D}_i]&=e_i^a\biggl[\left(-\fft18\mathcal D_\mu^{(-)}F_{\alpha\beta}^{(+)\,a}-\fft14P_\alpha^{(+-)\,ab}F_{\mu\beta}^{(-)\,b}-\fft18P_\mu^{(+-)\,ac}F_{\alpha\beta}^{(+)\,c}\right)\Gamma^{\alpha\beta}\nn\\
    &\qquad+\left(-\fft12\mathcal D_\mu^{(-)}P_\alpha^{(+-)\,ab}-\fft18F_{\mu\gamma}^{(-)\,b}F_{\gamma\alpha}^{(+)\,a}-\fft12P_\mu^{(+-)\,ac}P_\alpha^{(+-)\,cb}\right)\Gamma^\alpha\Gamma^b\nn\\
    &\qquad-\fft14P_\gamma^{(+-)\,ab}F_{\mu\gamma}^{(-)\,c}\Gamma^{bc}\biggr].
\end{align}
We then find
\begin{equation}
    \delta_\epsilon\psi_a'+T^{ab}\delta_\epsilon\psi_b^{(0)}-\fft1{16}F_{\mu\nu}^{(-)\,a}(F_{\mu\nu}^{(-)\,c}+F_{\mu\nu}^{(+)\,c})\delta_\epsilon\psi_c^{(0)}=\fft12P_\alpha^{(+-)\,ac}T^{cb}\Gamma^\alpha\Gamma^b\epsilon.
\end{equation}
Stripping off the $\delta_\epsilon$ and noting that $T^{ab}\delta_\epsilon\psi_b^{(0)}=e^{ai}\delta e_i^b\delta_\epsilon\psi_b^{(0)}$ suggests that the $\mathcal O(\alpha')$ shifted gaugino is given by
\begin{equation}
    \tilde\psi_i=\psi_i+e_i^a\left[-\fft14F_{\mu\nu}^{(-)\,a}\hat{\mathcal D}_\mu\psi_\nu-\fft12P_\mu^{(+-)\,ba}e_b^j(\hat{\mathcal D}_\mu\psi_j-\hat{\mathcal D}_j\psi_\mu)-\fft1{16}F_{\mu\nu}^{(-)\,a}(F_{\mu\nu}^{(-)\,b}+F_{\mu\nu}^{(+)\,b})e_b^j\psi_j\right].
\end{equation}
We then have
\begin{equation}
    \delta_\epsilon\tilde\psi_i=e_i^a\left[-\fft12P_\mu^{(+-)\,ab}(\delta^{bc}-\alpha'T^{bc})\Gamma^\mu\Gamma^c-\fft18F_{\mu\nu}^{(+)\,a}\Gamma^{\mu\nu}\right]\epsilon,
\label{eq:deltagau}
\end{equation}
where $T^{bc}$ is given in (\ref{eq:Tab}) and has both indices transforming in $O(d)_-$.  Note that this is the complete gaugino variation, including the leading order and $\mathcal O(\alpha')$ terms.

\subsection{Truncating the vector multiplets}

The reduction of heterotic supergravity on $T^d$ gives rise to a half-maximal supergravity in $10-d$ dimensions coupled to $d$ vector multiplets.  As demonstrated in \cite{Liu:2023fqq}, this theory admits a further truncation to pure half-maximal supergravity.  Since the vector multiplet consists of a gauge field, $d$ scalars, and corresponding gauginos, the truncation to pure supergravity involves removing the vector field strengths $F^{(+)\,a}$, scalar kinetic terms $P^{(+-)\,ab}$, and gauginos $\tilde\psi_i$.  This is accomplished by setting
\begin{equation}
    g_{ij}=\delta_{ij},\qquad b_{ij}=0,\qquad A_\mu^i=\fft12A_\mu^{(-)\,i},\qquad B_{\mu i}=-\fft12A_\mu^{(-)\,i},
\end{equation}
in the reduction ansatz, (\ref{eq:redans}).  With this choice, we find a trivial generalized vielbein, (\ref{eq:genviel})
\begin{equation}
    V=\begin{pmatrix}
        \mathbf1&-\mathbf1\\\mathbf1&\mathbf1
    \end{pmatrix},
\end{equation}
and resulting field strengths
\begin{equation}
    F^{(-)\,a}=dA^{(-)\,a},\qquad F^{(+)\,a}=0,\qquad P^{(+-)\,ab}=0.
\end{equation}
Note that the $\mathcal O(\alpha')$ shift of the internal metric
\begin{equation}
    \delta g_{ij}=\fft1{16}F_{\mu\nu}^{(-)\,i}F_{\mu\nu}^{(-)\,j},
\end{equation}
obtained from (\ref{eq:Tabshift}) and (\ref{eq:Tab}) is still present in the truncated reduction.

Making this truncation in the $O(d)_-\times O(d)_+$ Lagrangian, (\ref{eq:Lagd2}) and (\ref{eq:lag4d123}), gives
\begin{align}
    e^{-1}\mathcal L_{\partial^2}
    &=e^{-2\varphi}\biggl[R+4(\partial_\mu\varphi)^2-\fft1{12}(\tilde h_{\mu\nu\rho})^2-\fft18(F_{\mu\nu}^{(-)\,a})^2\nn\\
    &\kern3em+\fft{\alpha'}8\biggl((R_{\alpha\beta\gamma\delta}(\omega_+))^2-R_{\alpha\beta\gamma\delta}(\omega_+)F_{\alpha\beta}^{(-)\,a}F_{\gamma\delta}^{(-)\,a}+\fft12(\nabla_\gamma^{(+)}F_{\alpha\beta}^{(-)\,a})^2\nn\\
    &\kern6em+\fft18F_{\alpha\beta}^{(-)\,a}F_{\beta\gamma}^{(-)\,a}F_{\gamma\delta}^{(-)\,b}F_{\delta\alpha}^{(-)\,b}-\fft18F_{\alpha\beta}^{(-)\,a}F_{\beta\gamma}^{(-)\,b}F_{\gamma\delta}^{(-)\,a}F_{\delta\alpha}^{(-)\,b}\nn\\
    &\kern6em+\fft18F_{\alpha\beta}^{(-)\,a}F_{\alpha\beta}^{(-)\,b}F_{\gamma\delta}^{(-)\,a}F_{\gamma\delta}^{(-)\,b}\biggr)\biggr],
\end{align}
where $\tilde h=h-(\alpha'/4)\omega_{3L}^{(+)}$.  Furthermore, the supersymmetry variations, (\ref{eq:losusy}), (\ref{eq:deltagrav}) and (\ref{eq:deltadil}) truncate to
\begin{align}
    \delta_\epsilon\psi_\mu&=\biggl[\nabla_\mu(\omega_-)+\fft14F_{\mu\nu}^{(-)\,a}\Gamma^\nu\Gamma^a+\fft{\alpha'}{32}\biggl(2\Bigl(R_{\mu\alpha\gamma\delta}(\omega_+)F_{\gamma\delta}^{(-)\,b}-\fft18F_{\delta\epsilon}^{(-)\,b}F_{\delta\epsilon}^{(-)\,d}F_{\mu\alpha}^{(-)\,d}\Bigr)\Gamma^\alpha\Gamma^b\nn\\
    &\kern16em-\fft12F_{\delta\epsilon}^{(-)\,b}\nabla_\mu^{(+)}F_{\delta\epsilon}^{(-)\,a}\Gamma^{ab}\biggr)\biggr]\epsilon,\nn\\
    \delta\tilde\lambda&=\biggl[\Gamma^\mu\partial_\mu\varphi-\fft1{12}\tilde h_{\mu\nu\rho}\Gamma^{\mu\nu\rho}+\fft18F_{\mu\nu}^{(-)\,a}\Gamma^{\mu\nu}\Gamma^a\nn\\
    &\kern12em+\fft{\alpha'}{48}\biggl(\fft32\Bigl(R_{\alpha\beta\gamma\delta}(\omega_+)F_{\gamma\delta}^{(-)\,c}-\fft18F_{\delta\epsilon}^{(-)\,c}F_{\delta\epsilon}^{(-)\,d}F_{\alpha\beta}^{(-)\,d}\Bigr)\Gamma^{\alpha\beta}\Gamma^c\nn\\
    &\kern16em+\fft14F_{\alpha\beta}^{(-)\,a}F_{\beta\gamma}^{(-)\,b}F_{\gamma\alpha}^{(-)\,c}\Gamma^{abc}\biggr)\biggr]\epsilon,
\end{align}
while the gaugino variation, (\ref{eq:deltagau}), becomes trivial, $\delta_\epsilon\tilde\psi_i=0$.  Note that, in the above, the torsionful connection is given by $\omega_-=\omega-\fft12\tilde h$.  The truncated Lagrangian and supersymmetry variations agree with the previous results of \cite{Liu:2023fqq}.


\section{Field redefinitions and removing explicit derivatives}
\label{sec:redefine}

The complete reduced heterotic Lagrangian is given by $\mathcal L_{\mathrm{het}}=\mathcal L_{\partial^2}+\fft18\alpha'(\mathcal L_1+\mathcal L_2+\mathcal L_3)$ where $\mathcal L_{\partial^2}$ is given in (\ref{eq:Lagd2}) and the individual four-derivative pieces, $\mathcal L_n$, are given in (\ref{eq:L_1}), (\ref{eq:L_2}) and (\ref{eq:L_3}).  To facilitate comparison with the literature and to present a more canonical set of couplings, we can remove the explicit derivative terms in the four-derivative Lagrangian by a set of suitable field redefinitions.  This is performed by integration by parts along with the use of the leading order equations of motion, (\ref{eq:loeom}).

In particular, the derivative terms that we wish to remove are $(\mathcal D_\gamma^{(+)}F_{\alpha\beta}^{(-)\,a})^2$ from $\mathcal L_1$, $(\mathcal D_{[\mu}'^{(+)}F_{\nu]\gamma}^{(+)\,a})^2$ and $(\mathcal D_\gamma^{(+)}P_\alpha^{(+-)\,bd})^2$ from $\mathcal L_2$ and the $\mathcal D_\mu'$ derivative term in the first line of $\mathcal L_3$ given in (\ref{eq:L_3}) along with the three terms of $\mathcal L_3$ that involve a single explicit derivative.  The manipulation of these derivative terms is straightforward, although somewhat tedious.  As an example, consider the derivative term $(\mathcal D_\gamma^{(+)}F_{\alpha\beta}^{(-)\,a})^2$ in $\mathcal L_1$.  We start by explicitly writing out the torsionful covariant derivative $\mathcal D^{(+)}=\mathcal D+\fft12h$, and then using the Bianchi identity on $F_{\alpha\beta}^{(-)\,a}$
\begin{equation}
    \mathcal D_{[\alpha}F_{\beta\gamma]}^{(-)\,a}=P_{[\alpha}^{(+-)\,ba}F_{\beta\gamma]}^{(+)\,b},
\end{equation}
to obtain
\begin{align}
    (\mathcal D_\gamma^{(+)}F_{\alpha\beta}^{(-)\,a})^2&=-2\mathcal D_\mu F_{\alpha\beta}^{(-)\,a}\mathcal D_\alpha F_{\beta\mu}^{(-)\,a}+3P_{[\mu}^{(+-)\,ba}F_{\alpha\beta]}^{(+)\,b}P_\mu^{(+-)\,ca}F_{\alpha\beta}^{(+)\,c}\nn\\
    &\quad+2h_{\mu\nu\lambda}F_{\mu\beta}^{(-)\,a}\mathcal D_\nu F_{\lambda\beta}^{(-)\,a}+h_{\mu\nu\lambda}F_{\mu\beta}^{(-)\,a}h_{\nu\delta[\lambda}F_{\beta]\delta}^{(-)\,a}.
\end{align}
The first term can now be integrated by parts, keeping in mind the tree-level $e^{-2\varphi}$ factor.  This gives rise to a term of the form $\mathcal D_\mu\mathcal D_\alpha F_{\beta\mu}^{(-)\,a}$.  We then commute the covariant derivatives, which results in a Riemann term as well as a Ricci term that can be eliminated by use of the two-derivative Einstein equation.  We are then left with $\mathcal D_\mu F_{\beta\mu}^{(-)\,a}$, which can be rewritten using the two-derivative equation of motion for $F^{(-)}$.  After collecting terms and simplifying, we end up with the expression (\ref{eq:DFDFibp}) in Appendix~\ref{app:IBP}.  The remaining derivative terms are manipulated in a similar manner, and the final expressions are summarized in Appendix~\ref{app:IBP}.

After eliminating the explicit derivative terms, we are left with total derivatives and terms proportional to the two-derivative equations of motion.  The former do not affect the bulk equations of motion, so we just drop them, while the equation of motion terms can be eliminated by field redefinitions.  In this case, we end up with the final form of the $O(d)_-\times O(d)_+$ invariant reduced Lagrangian
\begin{equation}
    \mathcal L_{O(d)_-\times O(d)_+}=\mathcal L_{\partial^2}+\fft{\alpha'}8(\mathcal L_1+\mathcal L_2+\mathcal L_3),
\label{eq:fredlag}
\end{equation}
where $\mathcal L_{\partial^2}$ is given in (\ref{eq:Lagd2}), and
\begin{align}
    e^{-1}\mathcal{L}_1 = e^{-2\varphi}&\;\biggl[(R_{\alpha\beta\gamma\delta}(\omega_+))^2 + \dfrac{1}{3} h_{\alpha\beta\gamma}\omega_{3L}(\omega_+)_{\alpha\beta\gamma} - \dfrac{1}{2} R_{\alpha\beta\gamma\delta}(\omega_+)F^{(-)\, a}_{\alpha\beta} F^{(-)\, a}_{\gamma\delta}\nn\\
    & - \dfrac{1}{8} F^{(-)\, a}_{\alpha\beta} F^{(-)\, a }_{\beta\gamma} F^{(-)\, b}_{\gamma\delta} F^{(-)\, b}_{\delta\alpha} + \dfrac{1}{8} F^{(-)\, a}_{\alpha\beta} F^{(-)\, b }_{\beta\gamma} F^{(-)\, a}_{\gamma\delta} F^{(-)\, b}_{\delta\alpha} \biggr],
\label{eq:fredL_1}
\end{align}
\begin{align}
    e^{-1}\mathcal{L}_2 = e^{-2\varphi}&\;\biggl[-\dfrac{1}{8} F^{(+)\, a}_{\alpha\beta} F^{(+)\, a}_{\gamma\delta} F^{(+)\, b}_{\alpha\beta} F^{(+)\, b}_{\gamma\delta} + \dfrac{1}{4} F^{(+)\, a}_{\alpha\beta} F^{(+)\, b }_{\beta\gamma} F^{(+)\, a}_{\gamma\delta} F^{(+)\, b}_{\delta\alpha}\nn \\
    &- \dfrac{1}{8} F^{(+)\, a}_{\alpha\beta} F^{(+)\, a }_{\beta\gamma} F^{(+)\, b}_{\gamma\delta} F^{(+)\, b}_{\delta\alpha} - \dfrac{1}{2} P_{\gamma}^{(+-)\, ac} P_\gamma^{(+-)\, bc} F^{(+)\, a}_{\alpha\beta} F^{(+)\, b}_{\alpha\beta}\nn\\
    & - P_{\alpha}^{(+-)\, ab} P_\gamma^{(+-)\, cb} F^{(+)\, c}_{\alpha\beta} F^{(+)\, a}_{\beta\gamma} - P_{\alpha}^{(+-)\, ab} P_\gamma^{(+-)\, cb} F^{(+)\, a}_{\alpha\beta} F^{(+)\, c}_{\beta\gamma}\nn \\
    & + P_{\alpha}^{(+-)\, bc} P_\beta^{(+-)\, bc} F^{(+)\, a}_{\alpha\gamma} F^{(+)\, a}_{\gamma\beta} + 2 P_{\alpha}^{(+-)\, ab} P_\beta^{(+-)\, cb} P_{\alpha}^{(+-)\, cd} P_\beta^{(+-)\, ad}\nn \\
    &+2 P_{\alpha}^{(+-)\, ab} P_\beta^{(+-)\, cb} P_{\beta}^{(+-)\, cd} P_\alpha^{(+-)\, ad} - 2 P_{\alpha}^{(+-)\, ab} P_\alpha^{(+-)\, cb} P_{\beta}^{(+-)\, cd} P_\beta^{(+-)\, ad}\nn\\
    &- 2 P_{\alpha}^{(+-)\, ab} P_\beta^{(+-)\, ab} P_{\alpha}^{(+-)\, cd} P_\beta^{(+-)\, cd} \biggr],
\label{eq:fredL_2}
\end{align}
and
\begin{align}
    e^{-1}\mathcal{L}_3 = e^{-2\varphi} &\; \biggl[\dfrac{1}{3} h_{\alpha\beta\gamma}\omega_3(-Q^{(++)})_{\alpha\beta\gamma} + \dfrac{1}{8} F^{(-)\, a}_{\alpha\beta} F^{(-)\, a }_{\gamma\delta} F^{(+)\, b }_{\alpha\beta} F^{(+)\, a}_{\gamma\delta}\nn \\
    & -\dfrac{1}{4} F^{(-)\, a}_{\alpha\beta} F^{(+)\, b }_{\beta\gamma} F^{(-)\, a}_{\gamma\delta} F^{(+)\, b}_{\delta\alpha} - \dfrac{1}{4} F^{(-)\, a}_{\alpha\beta} F^{(-)\, a }_{\beta\gamma} F^{(+)\, b}_{\gamma\delta} F^{(+)\, b}_{\delta\alpha}\nn\\
    &+ \dfrac{1}{2} P_{\gamma}^{(+-)\, ab} P_\gamma^{(+-)\, ac} F^{(-)\, b}_{\alpha\beta} F^{(-)\, c}_{\alpha\beta} + 
     P_{\alpha}^{(+-)\, bc} P_\gamma^{(+-)\, bc} F^{(-)\, a}_{\alpha\beta} F^{(-)\, a}_{\beta\gamma}\nn\\ 
     &+ P_{\alpha}^{(+-)\, ba} P_\gamma^{(+-)\, bc} F^{(-)\, a}_{\alpha\beta} F^{(-)\, c}_{\beta\gamma}- P_{\alpha}^{(+-)\, bc} P_\gamma^{(+-)\, ba} F^{(-)\, a}_{\alpha\beta} F^{(-)\, c}_{\beta\gamma}\nn\\
     & + \dfrac{1}{2} h_{\alpha\beta\gamma} F^{(+)\,a}_{\alpha\delta} F^{(-)\, b}_{\beta\gamma} P_{\delta}^{(+-)\, ab} + \dfrac{1}{2} h_{\alpha\beta\gamma} F^{(-)\,b}_{\alpha\delta} F^{(+)\, a}_{\beta\gamma} P_{\delta}^{(+-)\, ab}\nn\\
     &  - h_{\alpha\beta\gamma} F^{(+)\,a}_{\alpha\delta} F^{(-)\, b}_{\gamma\delta} P_\beta^{(+-)\, ab}\biggr].
\label{eq:fredL_3}
\end{align}
Note that the $h$-terms in $\mathcal L_1$ and the first line of $\mathcal L_3$ combine with the leading order Lagrangian to form the $\alpha'$-corrected three-form
\begin{equation}
    \tilde h=h-\fft{\alpha'}4(\omega_{3L}(\omega_+)+\omega_3(-Q^{(++)})),
\end{equation}
with Bianchi identity
\begin{align}
    d\tilde h&=-\fft14\left(F^{(+)\,a}\wedge F^{(+)\,a}-F^{(-)\,a}\wedge F^{(-)\,a}\right)\nn\\
    &\quad-\fft{\alpha'}4\left(\Tr R(\omega_+)\wedge R(\omega_+)+P^{(+-)\,ab}\wedge P^{(+-)\,cb}\wedge P^{(+-)\,cd}\wedge P^{(+-)\,ad}\right).
\label{eq:hbian}
\end{align}

In order to arrive at this final form of the Lagrangian, we had to make use of the torsionful Riemann identities
\begin{align}
    \left(R_{\alpha\beta\gamma\delta}(\omega_-)-R_{\alpha\beta\gamma\delta}(\omega_+)\right)F_{\alpha\beta}^{(-)\,a}F_{\gamma\delta}^{(-)\,a}\kern-8em&\nn\\
    &=\dfrac14  F^{(+)\,a}_{\alpha\beta} F^{(-)\,b}_{\alpha\beta}F^{(+)\, a}_{\gamma\delta}  F^{(-)\, b}_{\gamma\delta} - \dfrac14  F^{(-)\,a}_{\alpha\beta}  F^{(-)\,b}_{\alpha\beta}  F^{(-)\, a}_{\gamma\delta}F^{(-)\, b}_{\gamma\delta}\nn \\
    &\quad - \dfrac12  F^{(+)\,a}_{\alpha\beta} F^{(-)\,b}_{\beta\gamma}F^{(+)\, a}_{\gamma\delta}  F^{(-)\, b}_{\delta\alpha} +  \dfrac12  F^{(-)\,a}_{\alpha\beta}  F^{(-)\,b}_{\beta\gamma} F^{(-)\, a}_{\gamma\delta}F^{(-)\, b}_{\delta\alpha},
\end{align}
and
\begin{align}
    \left(\dfrac{1}{2} R_{\alpha\beta\gamma\delta} (\omega_+) - R_{\alpha\gamma\beta\delta} (\omega_+)\right) F^{(+)\,a}_{\alpha\beta} F^{(+)\, a}_{\gamma\delta}\kern-8em&\nn\\
    &= \dfrac{1}{4}  h_{\alpha\gamma\epsilon} h_{\beta\delta\epsilon} F^{(+)\,a}_{\alpha\gamma} F^{(+)\, a}_{\beta\delta} -\dfrac{1}{2} h_{\alpha\gamma\epsilon} h_{\beta\delta\epsilon} F^{(+)\,a}_{\alpha\beta} F^{(+)\, a}_{\gamma\delta} \nn \\
    &\quad + \dfrac{3}{16}  F^{(+)\,a}_{\alpha\beta} F^{(-)\,b}_{\alpha\beta}F^{(+)\, a}_{\gamma\delta}  F^{(-)\, b}_{\gamma\delta} - \dfrac{3}{16}  F^{(+)\,a}_{\alpha\beta}  F^{(+)\,b}_{\alpha\beta}  F^{(+)\, a}_{\gamma\delta}F^{(+)\, b}_{\gamma\delta}\nn \\
    &\quad - \dfrac{3}{8}  F^{(+)\,a}_{\alpha\beta} F^{(-)\,b}_{\beta\gamma}F^{(+)\, a}_{\gamma\delta}  F^{(-)\, b}_{\delta\alpha} +  \dfrac{3}{8}  F^{(+)\,a}_{\alpha\beta}  F^{(+)\,b}_{\beta\gamma} F^{(+)\, a}_{\gamma\delta}F^{(+)\, b}_{\delta\alpha}.
\end{align}
These expressions can be obtained from the standard symmetries of torsion-free Riemann along with the leading order part of the $h$-Bianchi identity, (\ref{eq:hbian}).

We can also truncate the field redefined Lagrangian, (\ref{eq:fredlag}), to obtain the somewhat simpler form of the four-derivative corrected supergravity Lagrangian
\begin{align}
    e^{-1}\mathcal L_{\partial^2}
    &=e^{-2\varphi}\biggl[R+4(\partial_\mu\varphi)^2-\fft1{12}(\tilde h_{\mu\nu\rho})^2-\fft18(F_{\mu\nu}^{(-)\,a})^2\nn\\
    &\kern3em+\fft{\alpha'}8\biggl((R_{\alpha\beta\gamma\delta}(\omega_+))^2  - \dfrac{1}{2} R_{\alpha\beta\gamma\delta}(\omega_+)F^{(-)\, a}_{\alpha\beta} F^{(-)\, a}_{\gamma\delta}\nn\\
    &\kern6em - \dfrac{1}{8} F^{(-)\, a}_{\alpha\beta} F^{(-)\, a }_{\beta\gamma} F^{(-)\, b}_{\gamma\delta} F^{(-)\, b}_{\delta\alpha} + \dfrac{1}{8} F^{(-)\, a}_{\alpha\beta} F^{(-)\, b }_{\beta\gamma} F^{(-)\, a}_{\gamma\delta} F^{(-)\, b}_{\delta\alpha}\biggr)\biggr].
\end{align}
This agrees with the corresponding field redefined Lagrangian of \cite{Liu:2023fqq}.

\subsection{The Lagrangian with a torsion-free connection}

We have retained the torsionful connection, $\omega_+$, after dimensional reduction since the ten-dimensional heterotic theory is naturally formulated in terms of the ten-dimensional connection $\Omega_+$ \cite{Bergshoeff:1988nn,Bergshoeff:1989de}.  However, if desired, one can reformulate the $\mathcal O(\alpha')$ Lagrangian entirely in terms of the torsion-free connection $\omega$.

Note that the torsionful terms are entirely contained in the first line of $\mathcal L_1$ in (\ref{eq:fredL_1}).  By expanding out the torsion connection, integration by parts and use of the leading order equations of motion, we find
\begin{align}
    (R_{\alpha\beta\gamma\delta}(\omega_+))^2 = &\;\; (R_{\alpha\beta\gamma\delta})^2 + \dfrac{1}{2} R_{\alpha\beta\gamma\delta} h_{\alpha\beta\epsilon} h_{\gamma\delta\epsilon} - \dfrac{1}{8} (h^2_{\alpha\beta})^2 - \dfrac{1}{8} h^4\nn\\
    & +\dfrac{3}{32} F_{\alpha\beta}^{(-)\, a} F_{\gamma\delta}^{(-)\, a}  F_{\alpha\beta}^{(-)\, b} F_{\gamma\delta}^{(-)\, b} -\dfrac{3}{16}  F_{\alpha\beta}^{(-)\, a} F_{\beta\gamma}^{(-)\, b} F_{\gamma\delta}^{(-)\, a}  F_{\delta\alpha}^{(-)\, b} \nn\\
    &+\dfrac{3}{32} F_{\alpha\beta}^{(+)\, a} F_{\gamma\delta}^{(+)\, a}  F_{\alpha\beta}^{(+)\, b} F_{\gamma\delta}^{(+)\, b} -\dfrac{3}{16}  F_{\alpha\beta}^{(+)\, a} F_{\beta\gamma}^{(+)\, b} F_{\gamma\delta}^{(+)\, a}  F_{\delta\alpha}^{(+)\, b} \nn\\
    &-\dfrac{3}{16} F_{\alpha\beta}^{(-)\, a} F_{\gamma\delta}^{(-)\, a}  F_{\alpha\beta}^{(+)\, b} F_{\gamma\delta}^{(+)\, b} + \dfrac{3}{8}  F_{\alpha\beta}^{(-)\, a} F_{\beta\gamma}^{(+)\, b} F_{\gamma\delta}^{(-)\, a}  F_{\delta\alpha}^{(+)\, b}\nn \\
    &- h^2_{\alpha\beta} P_{\alpha}^{(+-)\, ab} P_\beta^{(+-)\, ab} - \dfrac{1}{4} h^2_{\alpha\beta} ( F_{\alpha\gamma}^{(-)\, a}  F_{\beta\gamma}^{(-)\, a} +  F_{\alpha\gamma}^{(+)\, a}  F_{\beta\gamma}^{(+)\, a})\nn\\
    &+\dfrac{1}{8}h_{\alpha\beta\epsilon} h_{\gamma\delta\epsilon} F_{\alpha\beta}^{(-)\, a} F_{\gamma\delta}^{(-)\, a} - \dfrac{1}{8}h_{\alpha\beta\epsilon} h_{\gamma\delta\epsilon} F_{\alpha\beta}^{(+)\, a} F_{\gamma\delta}^{(+)\, a}\nn\\
    &-\dfrac{1}{4}h_{\alpha\beta\epsilon} h_{\gamma\delta\epsilon} F_{\alpha\gamma}^{(-)\, a} F_{\beta\delta}^{(-)\, a} + \dfrac{1}{4}h_{\alpha\beta\epsilon} h_{\gamma\delta\epsilon} F_{\alpha\gamma}^{(+)\, a} F_{\beta\delta}^{(+)\, a},
\end{align}
along with
\begin{align}
    h_{\alpha\beta\gamma} \omega_{3L}(\omega_+)_{\alpha\beta\gamma} = &\;\; h_{\alpha\beta\gamma} \omega_{3L}(\omega)_{\alpha\beta\gamma} - 3 h_{\alpha\beta\epsilon}h_{\gamma\delta\epsilon} R_{\alpha\beta\gamma\delta} + \dfrac{1}{2} h^4\nn\\
    &- \dfrac{3}{16} h_{\alpha\beta\epsilon} h_{\gamma\delta\epsilon} F_{\alpha\beta}^{(-)\, a} F_{\gamma\delta}^{(-)\, a} + \dfrac{3}{16}h_{\alpha\beta\epsilon} h_{\gamma\delta\epsilon} F_{\alpha\beta}^{(+)\, a} F_{\gamma\delta}^{(+)\, a}\nn\\
    &+\dfrac{3}{8}h_{\alpha\beta\epsilon} h_{\gamma\delta\epsilon} F_{\alpha\gamma}^{(-)\, a} F_{\beta\delta}^{(-)\, a} - \dfrac{3}{8}h_{\alpha\beta\epsilon} h_{\gamma\delta\epsilon} F_{\alpha\gamma}^{(+)\, a} F_{\beta\delta}^{(+)\, a},
\end{align}
and
\begin{align}
    R(\omega_+)_{\alpha\beta\gamma\delta} F_{\alpha\beta}^{(-)\, a} F_{\gamma\delta}^{(-)\, a} = &\;\; R_{\alpha\beta\gamma\delta} F_{\alpha\beta}^{(-)\, a} F_{\gamma\delta}^{(-)\, a} - \dfrac{1}{2} h_{\alpha\beta\epsilon}h_{\gamma\delta\epsilon} F_{\alpha\gamma}^{(-)\, a} F_{\beta\delta}^{(-)\, a}\nn\\
    &+\dfrac{1}{8} F_{\alpha\beta}^{(-)\, a} F_{\gamma\delta}^{(-)\, a} F_{\alpha\beta}^{(-)\, b} F_{\gamma\delta}^{(-)\, b} - \dfrac{1}{8} F_{\alpha\beta}^{(-)\, a} F_{\gamma\delta}^{(-)\, a} F_{\alpha\beta}^{(+)\, b} F_{\gamma\delta}^{(+)\, b}\nn\\
    &-\dfrac{1}{4} F_{\alpha\beta}^{(-)\, a} F_{\beta\gamma}^{(-)\, b} F_{\gamma\delta}^{(-)\, a} F_{\delta\alpha}^{(-)\, b}  + \dfrac{1}{4} F_{\alpha\beta}^{(-)\, a} F_{\beta\gamma}^{(+)\, b}F_{\gamma\delta}^{(-)\, a} F_{\delta\alpha}^{(+)\, b} .
\end{align}
Here we have defined $h^2_{\alpha\beta}=h_{\alpha\gamma\delta}h_\beta{}^{\gamma\delta}$ and $h^4 = h_{\alpha\beta\gamma}h_\alpha{}^{\epsilon\delta}h_\beta{}^{\delta\eta}h_\gamma{}^{\eta\epsilon}$.  It is worth keeping in mind that these expressions made use of the equations of motion, corresponding to an implicit choice of field redefinitions.

With the above expansions, the reduced heterotic Lagrangian, (\ref{eq:fredlag}), can be rewritten with a torsion-free connection as
\begin{align}
    e^{-1}\mathcal L
    &=e^{-2\varphi}\biggl[R+4(\partial_\mu\varphi)^2-\fft1{12}(\tilde h_{\mu\nu\rho})^2-\fft18(F_{\mu\nu}^{(-)\,a})^2-\fft18(F_{\mu\nu}^{(+)\,a})^2-P_\mu^{(+-)\,ab}P_\mu^{(+-)\,ab}\nn\\
    &\kern4em+\fft{\alpha'}8\bigl(\mathcal L_1+\mathcal L_2+\mathcal L_3\bigr)\biggr],
\label{eq:tf4dlag}
\end{align}
where
\begin{align}
    e^{-1}\mathcal{L}_1 = e^{-2\varphi}&\;\biggl[(R_{\alpha\beta\gamma\delta})^2  - \dfrac{1}{2} R_{\alpha\beta\gamma\delta}F^{(-)\, a}_{\alpha\beta} F^{(-)\, a}_{\gamma\delta} - \dfrac{1}{8} (h_{\alpha\beta}^2)^2 \nn\\
    &+\dfrac{1}{24} h^4 - \dfrac{1}{2} R_{\alpha\beta\gamma\delta}h_{\alpha\beta\epsilon} h_{\gamma\delta\epsilon} + \dfrac{1}{32} F_{\alpha\beta}^{(-)\, a} F_{\gamma\delta}^{(-)\, a} F_{\alpha\beta}^{(-)\, b} F_{\gamma\delta}^{(-)\, b}\nn\\
    & - \dfrac{1}{8} F^{(-)\, a}_{\alpha\beta} F^{(-)\, a }_{\beta\gamma} F^{(-)\, b}_{\gamma\delta} F^{(-)\, b}_{\delta\alpha} + \dfrac{1}{16} F^{(-)\, a}_{\alpha\beta} F^{(-)\, b }_{\beta\gamma} F^{(-)\, a}_{\gamma\delta} F^{(-)\, b}_{\delta\alpha}\nn\\
    &+ \dfrac{1}{16} h_{\alpha\beta\epsilon}h_{\gamma\delta\epsilon} F^{(-)\, a}_{\alpha\beta}F^{(-)\, a}_{\gamma\delta} + \dfrac{1}{8} h_{\alpha\beta\epsilon}h_{\gamma\delta\epsilon} F^{(-)\, a}_{\alpha\gamma}F^{(-)\, a}_{\beta\delta}\nn \\
    &-\dfrac{1}{4} h^2_{\alpha\beta} F^{(-)\, a}_{\alpha\gamma}F^{(-)\, a}_{\beta\gamma}\biggr],
\end{align}
\begin{align}
    e^{-1}\mathcal{L}_2 = e^{-2\varphi}&\; \biggl[-\dfrac{1}{32} F^{(+)\, a}_{\alpha\beta} F^{(+)\, a}_{\gamma\delta} F^{(+)\, b}_{\alpha\beta} F^{(+)\, b}_{\gamma\delta} - \dfrac{1}{8} F^{(+)\, a}_{\alpha\beta} F^{(+)\, a }_{\beta\gamma} F^{(+)\, b}_{\gamma\delta} F^{(+)\, b}_{\delta\alpha}\nn \\
    & + \dfrac{1}{16} F^{(+)\, a}_{\alpha\beta} F^{(+)\, b }_{\beta\gamma} F^{(+)\, a}_{\gamma\delta} F^{(+)\, b}_{\delta\alpha} - \dfrac{1}{2} P_{\gamma}^{(+-)\, ac} P_\gamma^{(+-)\, bc} F^{(+)\, a}_{\alpha\beta} F^{(+)\, b}_{\alpha\beta}\nn\\
    & - P_{\alpha}^{(+-)\, ab} P_\gamma^{(+-)\, cb} F^{(+)\, a}_{\alpha\beta} F^{(+)\, c}_{\beta\gamma}- P_{\alpha}^{(+-)\, ab} P_\gamma^{(+-)\, cb} F^{(+)\, c}_{\alpha\beta} F^{(+)\, a}_{\beta\gamma} \nn \\
    & + P_{\alpha}^{(+-)\, bc} P_\beta^{(+-)\, bc} F^{(+)\, a}_{\alpha\gamma} F^{(+)\, a}_{\gamma\beta} + 2 P_{\alpha}^{(+-)\, ab} P_\beta^{(+-)\, cb} P_{\alpha}^{(+-)\, cd} P_\beta^{(+-)\, ad}\nn \\
    &+2 P_{\alpha}^{(+-)\, ab} P_\beta^{(+-)\, cb} P_{\beta}^{(+-)\, cd} P_\alpha^{(+-)\, ad} - 2 P_{\alpha}^{(+-)\, ab} P_\alpha^{(+-)\, cb} P_{\beta}^{(+-)\, cd} P_\beta^{(+-)\, ad}\nn\\
    &- 2 P_{\alpha}^{(+-)\, ab} P_\beta^{(+-)\, ab} P_{\alpha}^{(+-)\, cd} P_\beta^{(+-)\, cd} \biggr],
\end{align}
\begin{align}
    e^{-1}\mathcal{L}_3 = e^{-2\varphi} &\; \biggl[ - \dfrac{1}{4} F^{(-)\, a}_{\alpha\beta} F^{(-)\, a }_{\beta\gamma} F^{(+)\, b}_{\gamma\delta} F^{(+)\, b}_{\delta\alpha} - h^2_{\alpha\beta} P_\alpha^{(+-)\, ab} P_\beta^{(+-)\, ab} - \dfrac{1}{4} h^2_{\alpha\beta} F^{(+)\, a}_{\alpha\gamma} F^{(+)\,a}_{\beta\gamma}\nn\\
    &- \dfrac{1}{16} h_{\alpha\beta\epsilon} h_{\gamma\delta\epsilon} F^{(+)\,a}_{\alpha\beta} F^{(+)\,a}_{\gamma\delta} + \dfrac{1}{8} h_{\alpha\beta\epsilon} h_{\gamma\delta\epsilon} F^{(+)\,a}_{\alpha\gamma} F^{(+)\,a}_{\beta\delta}\nn\\
    &+ \dfrac{1}{2} P_{\gamma}^{(+-)\, ab} P_\gamma^{(+-)\, ac} F^{(-)\, b}_{\alpha\beta} F^{(-)\, c}_{\alpha\beta} + 
     P_{\alpha}^{(+-)\, bc} P_\gamma^{(+-)\, bc} F^{(-)\, a}_{\alpha\beta} F^{(-)\, a}_{\beta\gamma}\nn\\ 
     &+ P_{\alpha}^{(+-)\, ba} P_\gamma^{(+-)\, bc} F^{(-)\, a}_{\alpha\beta} F^{(-)\, c}_{\beta\gamma}- P_{\alpha}^{(+-)\, bc} P_\gamma^{(+-)\, ba} F^{(-)\, a}_{\alpha\beta} F^{(-)\, c}_{\beta\gamma}\nn\\
     & + \dfrac{1}{2} h_{\alpha\beta\gamma} F^{(+)\,a}_{\alpha\delta} F^{(-)\, b}_{\beta\gamma} P_{\delta}^{(+-)\, ab} + \dfrac{1}{2} h_{\alpha\beta\gamma} F^{(-)\,b}_{\alpha\delta} F^{(+)\, a}_{\beta\gamma} P_{\delta}^{(+-)\, ab}\nn\\
     &  - h_{\alpha\beta\gamma} F^{(+)\,a}_{\alpha\delta} F^{(-)\, b}_{\gamma\delta} P_\beta^{(+-)\, ab}\biggr].
\end{align}
Here, $\tilde h$ is
\begin{equation}
    \tilde h_{\alpha\beta\gamma}=h_{\alpha\beta\gamma}-\fft{\alpha'}4\left(\omega_{3L}(\omega)_{\alpha\beta\gamma}+\omega_3(-Q^{(++)})_{\alpha\beta\gamma}\right)
\end{equation}
This form of the $\mathcal O(\alpha')$ Lagrangian has the advantage that all dependence on the three-form $h$ is made explicit.

\section{Discussion}
\label{sec:discuss}

At the two-derivative level, Kaluza-Klein reduction on a torus is well established, and truncation to zero modes on the torus ensures that such a reduction remains consistent at all orders in the derivative expansion.  Nevertheless, we were able to gain further insight in the structure of supergravity reduction by examining the fermionic supersymmetries and focusing on the $O(d)_-\times O(d)_+$ symmetry under which the fermions transform.  With our conventions, $O(d)_-$ is the $R$-symmetry of the reduced supergravity, while $O(d)_+$ becomes a flavor symmetry.  Removal of the lower dimensional vector multiplets corresponds to truncating to $O(d)_+$ singlets.  Consistency of this truncation at $\mathcal O(\alpha')$ was demonstrated in \cite{Liu:2023fqq}, and the present results reduce to that case upon truncation.

The structure of the higher derivative corrections also reveals information about $T$-duality invariance of the heterotic Lagrangian.  At the massless supergravity level, the $T$-duality group $O(d,d;\mathbb Z)$ extends to the continuous symmetry $O(d,d;\mathbb R)$, and the $O(d,d;\mathbb R)$ invariant reduced Lagrangian was explicitly constructed in \cite{Eloy:2020dko}.  Naturally one would expect the $O(d)_-\times O(d)_+$ invariant Lagrangian, (\ref{eq:tf4dlag}), to agree with the $O(d,d;\mathbb R)$ invariant Lagrangian of \cite{Eloy:2020dko}.  In order to make this comparison, we must convert between the top and bottom of the $O(d,d;\mathbb R)/O(d)_-\times O(d)_+$ coset using the generalized vielbein, $V$, given explicitly in (\ref{eq:genviel}).  In a matrix notation, the scalar matrix $\mathcal S_M{}^N$ of \cite{Eloy:2020dko} is related to $V^A{}_N$ according to $\mathcal S=\mathcal H\eta=\fft12V^TV\eta$.  From this, we can deduce various relations such as
\begin{align}
    \mathcal F_{\alpha\beta}\eta\mathcal F_{\gamma\delta}&=\fft12\left(F_{\alpha\beta}^{(+)\,a}F_{\gamma\delta}^{(+)\,a}-F_{\alpha\beta}^{(-)\,a}F_{\gamma\delta}^{(-)\,a}\right),\nn\\
    \mathcal F_{\alpha\beta}\mathcal H\mathcal F_{\gamma\delta}&=\fft12\left(F_{\alpha\beta}^{(+)\,a}F_{\gamma\delta}^{(+)\,a}+F_{\alpha\beta}^{(-)\,a}F_{\gamma\delta}^{(-)\,a}\right),\nn\\
    \partial_\mu\mathcal S&=V^T\begin{pmatrix}0&P_\mu^{(-+)}\\P_\mu^{(+-)}&0\end{pmatrix}V\eta.
\end{align}
Making use of these relations, along with the identity $V\eta V^T=-2\Pi$, where
\begin{equation}
    \Pi=\begin{pmatrix}\mathbf1&0\\0&-\mathbf1\end{pmatrix},
\end{equation}
we find that the $O(d)_-\times O(d)_+$ invariant Lagrangian, (\ref{eq:tf4dlag}) indeed agrees with the result of \cite{Eloy:2020dko} up to a sign convention of taking $H_{MNP}\to-H_{MNP}$.

Because of the distinct roles played by $O(d)_-$ and $O(d)_+$ as $R$-symmetry and flavor symmetry, respectively, none of the forms of the reduced Lagrangian presented above exhibit symmetry under the interchange of these two $O(d)$ factors.  Such an interchange would be equivalent to the world-sheet parity transformation, $H\to-H$, which is not present for the heterotic string.

Since the bosonic string is invariant under world-sheet parity, had we reduced the bosonic string Lagrangian, we would have found it to be symmetric under interchange of $O(d)_-$ and $O(d)_+$.  The bosonic string reduction was also considered in \cite{Eloy:2020dko}, and the reduced Lagrangian takes a simpler form than the heterotic case, as there is no Lorentz Chern-Simons form to consider.   World-sheet parity imposes a further constraint on the possible terms that can show up in the reduced action.  For example, the fields $\mathcal F$ and $\mathcal S$ transform under world-sheet parity as
\begin{equation}
    \mathcal F\to\Pi\mathcal F,\qquad\mathcal H\to\Pi\mathcal H\Pi.
\end{equation}
We then see that the combination $\mathcal F_{\alpha\beta}\mathcal H\mathcal F_{\gamma\delta}$ is even under world-sheet parity while $\mathcal F_{\alpha\beta}\eta\mathcal F_{\gamma\delta}$ is odd.  Up to an overall normalization, the reduced bosonic string Lagrangian has a coupling of the form
\begin{equation}
    R_{\alpha\beta\gamma\delta}\mathcal F_{\alpha\beta}\mathcal H\mathcal F_{\gamma\delta}=\fft12R_{\alpha\beta\gamma\delta}\left(F_{\alpha\beta}^{(+)\,a}F_{\gamma\delta}^{(+)\,a}+F_{\alpha\beta}^{(-)\,a}F_{\gamma\delta}^{(-)\,a}\right),
\end{equation}
which is symmetric under $F^{(+)}\leftrightarrow F^{(-)}$ interchange.  The heterotic string has this same term, but adds to it $-R_{\alpha\beta\gamma\delta}\mathcal F_{\alpha\beta}\eta\mathcal F_{\gamma\delta}$ (in our sign convention for $H_{MNP}$) which originates from the ten-dimensional Lorentz Chern-Simons term.  The result is then
\begin{equation}
    R_{\alpha\beta\gamma\delta}\mathcal F_{\alpha\beta}(\mathcal H-\eta)\mathcal F_{\gamma\delta}=R_{\alpha\beta\gamma\delta}F_{\alpha\beta}^{(-)\,a}F_{\gamma\delta}^{(-)\,a},
\end{equation}
which singles out $F^{(-)}$ as the graviphotons.

Clearly, the combination of T-duality invariance and world-sheet parity is more restrictive than T-duality alone.  Since the type II strings are invariant under world-sheet parity (along with space-time parity in the IIB case), it would be interesting to see if this combination of symmetries could shed further light on the structure of T-duality invariant couplings of the type II string from an $O(d)_-\times O(d)_+$ point of view.  A full consideration would involve the Ramond-Ramond sector as well, in which case the $O(d)_-\times O(d)_+$ symmetry would be enhanced to the bottom of the U-duality group.  A natural way to pursue this investigation could be through the use of exceptional field theory \cite{Hohm:2013pua,Hohm:2013vpa,Hohm:2013uia,Hohm:2014fxa}.  Note that U-duality also arises in bosonic and heterotic compactifications to three dimensions, and the interplay between $\mathcal O(\alpha')$ corrections and U-duality has been investigated in \cite{Eloy:2022vsq} for this case.

Here we have only focused on heterotic supergravity.  Had we included the ten-dimensional heterotic gauge fields, we would have had a $O(d,d+16)/O(d)_-\times O(d+16)_+$ coset and a corresponding enlarged flavor symmetry.  Since we have written the reduced Lagrangian in a manifestly $O(d)_-\times O(d)_+$ notation, would be natural to simply extend all $O(d)_+$ indices to run from $1$ to $d+16$.  While additional surprises may arise, we conjecture that this is sufficient to arrive at the $O(d)_-\times O(d+16)_+$ form of the reduced heterotic Lagrangian.  This would be an example of symmetry leading the way to a more complete result with no additional work.

\acknowledgments

We wish to thank R.\ Saskowski for useful discussions.  This work is supported in part by the U.S. Department of Energy under grant DE-SC0007859.  SJ is supported in part by a Leinweber Graduate Summer Fellowship.

\appendix
\section{Connection, curvature and Chern-Simons form}
\label{app:Omega}

Here we present the reduction of the ten-dimensional torsionful connection, curvature and Chern-Simons form.  We start by noting that the reduction of the three-form $H$ takes the form
\begin{equation}
    H=\fft16h_{\mu\nu\rho}dx^\mu\wedge dx^\nu\wedge dx^\nu+\fft12(F^{(+)\,a}-F^{(-)\,a})\eta^a+\fft12(P^{(+-)\,ba}-P^{(+-)\,ab})\eta^a\wedge\eta^b,
\end{equation}
when written in terms of the $O(d)_+\times O(d)_-$ fields.  Here $\eta^a=e^a_i\eta^i$ where $\eta_i$ is given in (\ref{eq:redans}).  Writing this as a one-form $\mathcal H^{AB}=H_M{}^{AB}dx^M$ and combining with the Kaluza-Klein reduction of the torsion-free spin connection to form the torsionful connection, $\Omega_\pm=\Omega\pm\fft12\mathcal H$, we find
\begin{equation}
    \Omega_\pm=\begin{pmatrix}\omega_\pm^{\alpha\beta}-\fft12F_{\alpha\beta}^{(\mp)\,a}\eta^a&\fft12F_{\mu\alpha}^{(\mp)\,b}dx^\mu-P_\alpha^{(\pm\mp)\,bc}\eta^c\\-\fft12F_{\mu\beta}^{(\pm)\,a}dx^\mu+P_\beta^{(\pm\mp)\,ac}\eta^c&-Q^{(\pm\pm)\,ab}\end{pmatrix}.
\end{equation}

The torsionful Riemann tensor is computed as $R(\Omega_+)=d\Omega_++\Omega_+\wedge\Omega_+$.  We find for its components
\begin{align}
    R_{\gamma\delta}{}^{\alpha\beta}(\Omega_+)&=R_{\gamma\delta}{}^{\alpha\beta}(\omega_+)-\fft12F_{[\gamma|\alpha}^{(+)\,c}F_{\delta]\beta}^{(+)\,c}-\fft14F_{\alpha\beta}^{(-)\,c}F_{\gamma\delta}^{(-)\,c}-\textcolor{red}{\fft14F_{\alpha\beta}^{(-)\,c}F_{\gamma\delta}^{(+)\,c}},\nn\\
    R_{\gamma d}{}^{\alpha\beta}(\Omega_+)&=-\fft12\mathcal D_\gamma^{(+)\,c}F_{\alpha\beta}^{(-)\,c}+F_{\gamma[\alpha}^{(+)\,c}P_{\beta]}^{(+-)\,cd}-\textcolor{red}{\fft12P_\gamma^{(+-)\,cd}F_{\alpha\beta}^{(-)\,c}},\nn\\
    R_{cd}{}^{\alpha\beta}(\Omega_+)&=\fft12F_{\alpha\gamma}^{(+)\,[c}F_{\gamma\beta}^{(-)\,d]}-2P_\alpha^{(+-)e[c}P_\beta^{(+-)\,e|d]},\nn\\
    R_{\gamma\delta}{}^{\alpha b}(\Omega_+)&=\mathcal D_{[\gamma}^{\prime(+)}F_{\delta]\alpha}^{(+)\,b}-\fft12P_\alpha^{(+-)\,bc}F_{\gamma\delta}^{(-)\,c}-\textcolor{red}{\fft12P_\alpha^{(+-)\,bc}F_{\gamma\delta}^{(+)\,c}},\nn\\
    R_{\gamma d}{}^{\alpha b}(\Omega_+)&=\fft14F_{\alpha\epsilon}^{(-)\,d}F_{\gamma\epsilon}^{(+)\,b}-\mathcal D_\gamma^{(+)}P_\alpha^{(+-)\,bd}-\textcolor{red}{P_\alpha^{(+-)\,bc}P_\gamma^{(+-)\,cd}},\nn\\
    R_{cd}{}^{\alpha b}(\Omega_+)&=F_{\alpha\gamma}^{(-)\,[c}P_\gamma^{(+-)\,b|d]},\nn\\
    R_{\gamma\delta}{}^{ab}(\Omega_+)&=-\fft12F_{[\gamma|\epsilon}^{(+)\,a}F_{\delta]\epsilon}^{(+)\,b}-2P_{[\gamma}^{(+-)\,ac}P_{\delta]}^{(+-)\,bc},\nn\\
    R_{\gamma d}{}^{ab}(\Omega_+)&=F_{\gamma\epsilon}^{(+)\,[a}P_\epsilon^{(+-)\,b]d},\nn\\
    R_{cd}{}^{ab}(\Omega_+)&=-2P_\gamma^{(+-)\,[a|c}P_\gamma^{(+-)\,b]d}.
\label{eq:Riemann+}
\end{align}

Finally, we compute the torsionful Lorentz-Chern-Simons form
\begin{equation}
    \omega_{3L}(\Omega_+)=\Tr\left(\Omega_+\wedge d\Omega_++\fft23\Omega_+\wedge\Omega_+\wedge\Omega_+\right).
\end{equation}
A direct dimensional reduction of $\omega_{3L}(\Omega_+)$ gives rise to a non-lower-dimensional covariant expression.  However, the Lorentz-Chern-Simons form only shows up in combination with $H=dB$, we can absorb the non-covariant term with a shift in $B$.  To be explicit, we define the shifted torsionful Lorentz-Chern-Simons form
\begin{equation}
    \tilde\omega_{3L}(\Omega_+)=\omega_{3L}(\Omega_+)+d\left(\fft12\omega_+^{\alpha\beta}F_{\alpha\beta}^{(-)\,c}\eta^c+F_{\mu\alpha}^{(+)\,b}P_\alpha^{(+-)\,bd}dx^\mu\wedge\eta^d\right),
\end{equation}
where the exact term is to be removed by a corresponding shift in $B$.  Then, the components of $\tilde\omega_{3L}(\Omega_+)$ are
\begin{align}
    \tilde\omega_{3L}(\Omega_+)_{\alpha\beta\gamma}&=\omega_{3L}(\omega_+)_{\alpha\beta\gamma}+\omega_{3}(-Q^{(++)})_{\alpha\beta\gamma}+3F_{\epsilon[\alpha}^{(+)\,a}\mathcal D_\beta^{\prime(+)}F_{\gamma]\epsilon}^{(+)\,a},\nn\\
    \tilde\omega_{3L}(\Omega_+)_{\alpha\beta c}&=R_{\alpha\beta}{}^{\gamma\delta}(\omega_+)F_{\gamma\delta}^{(-)\,c}+4P_\delta^{(+-)\,dc}\mathcal D_{[\alpha}^{\prime(+)}F_{\beta]\delta}^{(+)\,d}-\fft18F_{\delta\epsilon}^{(-)\,c}\textcolor{red}{F_{\delta\epsilon}^{(-)\,d}F_{\alpha\beta}^{(+)\,d}}\nn\\
    &\quad-\fft18F_{\delta\epsilon}^{(-)\,c}F_{\delta\epsilon}^{(-)\,d}F_{\alpha\beta}^{(-)\,d}-P_\delta^{(+-)\,dc}\textcolor{red}{P_\delta^{(+-)\,de}F_{\alpha\beta}^{(+)\,e}}-P_\delta^{(+-)\,dc}P_\delta^{(+-)\,de}F_{\alpha\beta}^{(-)\,e}\nn\\
    &\quad+\fft12F_{\delta\epsilon}^{(-)\,c}F_{\alpha\epsilon}^{(+)\,d}F_{\beta\delta}^{(+)\,d},\nn\\
    \tilde\omega_{3L}(\Omega_+)_{\alpha bc}&=-4P_\delta^{(+-)\,d[c|}\mathcal D_\alpha^{(+)}P_\delta^{(+-)\,d|b]}-\fft12F_{\delta\epsilon}^{(-)\,[c}\mathcal D_\alpha^{(+)}F_{\delta\epsilon}^{(-)\,b]}-\fft12F_{\delta\epsilon}^{(-)\,[c|}\textcolor{red}{F_{\delta\epsilon}^{(-)\,d}P_\alpha^{(+-)\,d|b]}}\nn\\
    &\quad-2F_{\delta\epsilon}^{(-)\,[b|}F_{\alpha\delta}^{(+)\,d}P_\epsilon^{(+-)\,d|c]}-4P_\delta^{(+-)\,d[c|}\textcolor{red}{P_\delta^{(+-)\,de}P_\alpha^{(+-)\,e|b]}},\nn\\
    \tilde\omega_{3L}(\Omega_+)_{abc}&=-\fft12F_{\alpha\beta}^{(-)\,a}F_{\beta\gamma}^{(-)\,b}F_{\gamma\alpha}^{(-)\,c}+6F_{\delta\epsilon}^{(-)\,[a|}P_\epsilon^{(+-)\,d|b|}P_\delta^{(+-)\,d|c]}.
\label{eq:LCS+}
\end{align}

\section{Removal of the explicit derivative terms}
\label{app:IBP}

We can manipulate the four-derivative Lagrangian by integration by parts and use of the lowest order equations of motion to rewrite the derivative terms.  We find
\begin{align}
    \fft12(\mathcal D_\gamma^{(+)}F_{\alpha\beta}^{(-)\,a})^2&=\fft12R_{\alpha\beta\gamma\delta}(\omega_-)F_{\alpha\beta}^{(-)\,a}F_{\gamma\delta}^{(-)\,a}-\fft14F_{\alpha\beta}^{(-)\,a}F_{\beta\gamma}^{(-)\,a}(F_{\gamma\delta}^{(+)\,b}F_{\delta\alpha}^{(+)\,b}+F_{\gamma\delta}^{(-)\,b}F_{\delta\alpha}^{(-)\,b})\nn\\
    &\quad+P_\alpha^{(+-)\,ab}P_\gamma^{(+-)\,ab}F_{\alpha\beta}^{(-)\,c}F_{\beta\gamma}^{(-)\,c}-2P_{[\gamma}^{(+-)\,ca}P_{\alpha]}^{(+-)\,cb}F_{\alpha\beta}^{(-)\,a}F_{\beta\gamma}^{(-)\,b}\nn\\
    &\quad+P_\gamma^{(+-)\,ac}P_\delta^{(+-)\,bc}F_{\beta\gamma}^{(+)\,a}F_{\beta\delta}^{(+)\,b}+\fft12P_\mu^{(+-)\,ac}P_\mu^{(+-)\,bc}F_{\alpha\beta}^{(+)\,a}F_{\alpha\beta}^{(+)\,b}\nn\\
    &\quad+P_\gamma^{(+-)\,ca}P_\beta^{(+-)\,ba}F_{\alpha\beta}^{(+)\,c}F_{\gamma\alpha}^{(+)\,b}+\fft12h_{\beta\mu\nu}P_\gamma^{(+-)\,ab}F_{\beta\gamma}^{(+)\,a}F_{\mu\nu}^{(-)\,b}\nn\\
    &\quad-\fft12h_{\beta\gamma\delta}P_\alpha^{(+-)\,ca}F_{\gamma\delta}^{(+)\,c}F_{\alpha\beta}^{(-)\,a}-h_{\beta\gamma\delta}P_\gamma^{(+-)\,ca}F_{\delta\alpha}^{(+)\,c}F_{\alpha\beta}^{(-)\,a}\nn\\
    &\quad-e^{2\varphi}\nabla_\alpha(e^{-2\varphi}F_{\alpha\beta}^{(-)\,a}P_\gamma^{(+-)\,ca}F_{\beta\gamma}^{(+)\,c})-e^{2\varphi}\nabla_\mu(e^{-2\varphi}F_{\alpha\beta}^{(-)\,a}\mathcal D_\alpha F_{\beta\mu}^{(-)\,a})\nn\\
    &\quad+F_{\alpha\beta}^{(-)\,a}F_{\beta\gamma}^{(-)\,a}\mathcal E_{\alpha\gamma}+F_{\alpha\beta}^{(-)\,a}\mathcal D_\alpha\mathcal E_\beta^{(-)\,a}-P_\gamma^{(+-)\,ba}F_{\beta\gamma}^{(+)\,b}\mathcal E_\beta^{(-)\,a},
\label{eq:DFDFibp}
\end{align}
and
\begin{align}
    2\mathcal D_\mu^{'(+)}F_{\mu\alpha}^{(+)\,a}\mathcal D_{[\mu}^{'(+)}F_{\nu]\alpha}^{(+)\,a}
    &=\fft12R_{\alpha\beta\gamma\delta}(\omega_+)F_{\alpha\beta}^{(+)\,a}F_{\gamma\delta}^{(+)\,a}-\fft14F_{\alpha\beta}^{(+)\,a}F_{\beta\gamma}^{(+)\,a}(F_{\gamma\delta}^{(+)\,b}F_{\delta\alpha}^{(+)\,b}+F_{\gamma\delta}^{(-)\,b}F_{\delta\alpha}^{(-)\,b})\nn\\
    &\quad+P_\gamma^{(+-)\,ab}P_\delta^{(+-)\,ab}F_{\gamma\beta}^{(+)\,c}F_{\beta\delta}^{(+)\,c}-P_\delta^{(+-)\,ac}P_\gamma^{(+-)\,bc}F_{\gamma\beta}^{(+)\,a}F_{\beta\delta}^{(+)\,b}\nn\\
    &\quad+P_\delta^{(+-)\,ac}P_\gamma^{(+-)\,bc}F_{\gamma\beta}^{(+)\,b}F_{\beta\delta}^{(+)\,a}+P_\alpha^{(+-)\,ab}P_\alpha^{(+-)\,ac}F_{\mu\nu}^{(-)\,b}F_{\mu\nu}^{(-)\,c}\nn\\
    &\quad-P_\gamma^{(+-)\,ab}P_\delta^{(+-)\,ac}F_{\gamma\beta}^{(-)\,b}F_{\beta\delta}^{(-)\,c}+2P_\delta^{(+-)\,ab}P_\gamma^{(+-)\,ac}F_{\gamma\beta}^{(-)\,b}F_{\beta\delta}^{(-)\,c}\nn\\
    &\quad-\fft12h_{\beta\mu\nu}P_\delta^{(+-)\,ab}F_{\mu\nu}^{(+)\,a}F_{\beta\delta}^{(-)\,b}+\fft12h_{\beta\mu\nu}P_\delta^{(+-)\,ab}F_{\beta\delta}^{(+)\,a}F_{\mu\nu}^{(-)\,b}\nn\\
    &\quad+h_{\alpha\gamma\delta}P_\alpha^{(+-)\,ab}F_{\beta\delta}^{(+)\,a}F_{\beta\gamma}^{(-)\,b}\nn\\
    &\quad+e^{2\varphi}\nabla_\gamma(e^{-2\varphi}F_{\beta\gamma}^{(+)\,a}P_\delta^{(+-)\,ab}F_{\beta\delta}^{(-)\,b})-e^{2\varphi}\nabla_\alpha(e^{-2\varphi}F_{\beta\gamma}^{(+)\,a}\mathcal D_\gamma F_{\alpha\beta}^{(+)\,a})\nn\\
    &\quad+F_{\beta\gamma}^{(+)\,a}F_{\delta\beta}^{(+)\,a}\mathcal E_{\gamma\delta}-F_{\beta\gamma}^{(+)\,a}\mathcal D_\gamma\mathcal E_\beta^{(+)\,a}-P_\delta^{(+-)\,ab}F_{\beta\delta}^{(-)\,b}\mathcal E_\beta^{(+)\,a},
\end{align}
along with
\begin{align}
    4(\mathcal D_\alpha^{(+)}P_\beta^{(+-)\,ab})^2&=-P_\alpha^{(+-)\,ab}P_\beta^{(+-)\,ab}(F_{\alpha\gamma}^{(+)\,c}F_{\beta\gamma}^{(+)\,c}+F_{\alpha\gamma}^{(-)\,c}F_{\beta\gamma}^{(-)\,c})\nn\\
    &\quad-4P_\alpha^{(+-)\,ab}P_\beta^{(+-)\,ab}P_\alpha^{(+-)\,cd}P_\beta^{(+-)\,cd}+8P_\alpha^{(+-)\,ab}P_\beta^{(+-)\,cb}P_\alpha^{(+-)\,cd}P_\beta^{(+-)\,ad}\nn\\
    &\quad-4P_\alpha^{(+-)\,ab}P_\alpha^{(+-)\,cb}P_\beta^{(+-)\,cd}P_\beta^{(+-)\,ad}-4P_\alpha^{(+-)\,ab}P_\beta^{(+-)\,cb}P_\beta^{(+-)\,cd}P_\alpha^{(+-)\,ad}\nn\\
    &\quad+\fft14F_{\mu\nu}^{(+)\,a}F_{\mu\nu}^{(-)\,b}F_{\lambda\sigma}^{(+)\,a}F_{\lambda\sigma}^{(-)\,b}\nn\\
    &\quad+4e^{2\varphi}\nabla_\alpha(e^{-2\varphi}P_\beta^{(+-)\,ab}\mathcal D_\alpha P_\beta^{(+-)\,ab})-e^{2\varphi}\nabla_\beta(e^{-2\varphi}P_\alpha^{(+-)\,ab}F_{\mu\nu}^{(+)\,a}F_{\mu\nu}^{(-)\,b})\nn\\
    &\quad-4P_\alpha^{(+-)\,ab}P_\beta^{(+-)\,ab}\mathcal E_{\alpha\beta}-4P_\alpha^{(+-)\,ab}\mathcal D_\alpha\mathcal E^{(+-)\,ab}+F_{\mu\nu}^{(+)\,a}F_{\mu\nu}^{(-)\,b}\mathcal E^{(+-)\,ab},
\end{align}
and
\begin{align}
    h_{\alpha\beta\gamma}F_{\alpha\delta}^{(+)\, a} \mathcal{D}_\beta^{'(+)}F_{\gamma\delta}^{(+)\, a} =&\;\; \;\;\dfrac{1}{8}  F_{\alpha\beta}^{(+)\,a} F_{\gamma\delta}^{(+)\,a} ( F_{\alpha\gamma}^{(-)\,b} F_{\beta\delta}^{(-)\,b} -  F_{\alpha\gamma}^{(+)\,b} F_{\beta\delta}^{(+)\,b})\nn \\
    &-\dfrac{1}{16} F_{\alpha\beta}^{(+)\,a} F_{\gamma\delta}^{(+)\,a} (  F_{\alpha\beta}^{(-)\,b} F_{\gamma\delta}^{(-)\,b} -  F_{\alpha\beta}^{(+)\,b} F_{\gamma\delta}^{(+)\,b})\nn \\
    &+\dfrac{1}{4}h_{\alpha\beta\gamma} h_{\alpha\delta\epsilon} F_{\beta\gamma}^{(+)\, a} F_{\delta\epsilon}^{(+)\, a} - \dfrac{1}{2} h_{\beta\alpha\gamma}h_{\beta\delta\epsilon} F^{(+)\, a}_{\alpha\delta} F^{(+)\, a}_{\gamma \epsilon}\nn\\
    &+\dfrac{1}{2} h_{\alpha\beta\gamma} F^{(+)\,a}_{\alpha\delta} F^{(-)\, b}_{\beta\gamma} P_{\delta}^{(+-)\, ab} - \dfrac{1}{2} h_{\alpha\beta\gamma} F^{(-)\,b}_{\alpha\delta} F^{(+)\, a}_{\beta\gamma} P_{\delta}^{(+-)\, ab} \nn\\
    &+ h_{\alpha\beta\gamma} F^{(+)\,a}_{\alpha\delta} F^{(-)\, b}_{\gamma\delta} P_\beta^{(+-)\, ab}\nn\\
    &- \dfrac{1}{2}\, e^{2\phi} \nabla_\delta(e^{-2\phi} h_{\alpha\beta\gamma} F^{(+)\, a}_{\alpha\delta} F^{(+)\,a}_{\beta\gamma})-\dfrac{1}{2} \, h_{\alpha\beta\gamma} F^{(+)\, a}_{\beta\gamma} \mathcal{E}_\alpha^{(+)\, a}.
\end{align}
and
\begin{align}
    \kern6em&\kern-6em-4P_\gamma^{(+-)\,ab}F_{\mu\nu}^{(-)\,b}\mathcal D_{[\mu}^{'(+)}F_{\nu]\gamma}^{(+)\,a}-2P_\beta^{(+-)\,ad}F_{\gamma\alpha}^{(+)\,a}\mathcal D_\gamma^{(+)}F_{\alpha\beta}^{(-)\,d}-2F_{\alpha\beta}^{(-)\,d}F_{\gamma\beta}^{(+)\,b}\mathcal D_\gamma^{(+)}P_\alpha^{(+-)\,bd}\nn\\
    &=2P_\alpha^{(+-)\,ab}P_\gamma^{(+-)\,ac}F_{\alpha\beta}^{(-)\,b}F_{\beta\gamma}^{(-)\,c}-4P_\gamma^{(+-)\,ab}P_\alpha^{(+-)\,ac}F_{\alpha\beta}^{(-)\,b}F_{\beta\gamma}^{(-)\,c}\nn\\
    &\quad-2P_\mu^{(+-)\,ab}P_\mu^{(+-)\,ac}F_{\alpha\beta}^{(-)\,b}F_{\alpha\beta}^{(-)\,c}-4P_\alpha^{(+-)\,ab}P_\gamma^{(+-)\,cb}F_{\gamma\beta}^{(+)\,a}F_{\beta\alpha}^{(+)\,c}\nn\\
    &\quad-2P_\mu^{(+-)\,ab}P_\mu^{(+-)\,cb}F_{\alpha\beta}^{(+)\,a}F_{\alpha\beta}^{(+)\,c}-\fft12F_{\mu\nu}^{(+)\,a}F_{\mu\nu}^{(-)\,b}F_{\alpha\beta}^{(+)\,a}F_{\alpha\beta}^{(-)\,b}\nn\\
    &\quad+2e^{2\varphi}\nabla_\gamma(e^{-2\varphi}P_\alpha^{(+-)\,ab}F_{\beta\gamma}^{(+)\,a}F_{\alpha\beta}^{(-)\,b})+2e^{2\varphi}\nabla_\alpha(e^{-2\varphi}P_\alpha^{(+-)\,ab}F_{\mu\nu}^{(+)\,a}F_{\mu\nu}^{(-)\,b})\nn\\
    &\quad-2P_\alpha^{(+-)\,ab}F_{\alpha\beta}^{(-)\,b}\mathcal E_\beta^{(+)\,a}-2F_{\mu\nu}^{(+)\,a}F_{\mu\nu}^{(-)\,b}\mathcal E^{(+-)\,ab}.
\end{align}

The explicit derivative terms added together gives
\begin{align}
    e^{-1}\mathcal L_{\mathcal D}&=\fft12R_{\alpha\beta\gamma\delta}(\omega_-)F_{\alpha\beta}^{(-)\,a}F_{\gamma\delta}^{(-)\,a}-\fft14F_{\alpha\beta}^{(-)\,a}F_{\beta\gamma}^{(-)\,a}F_{\gamma\delta}^{(-)\,b}F_{\delta\alpha}^{(-)\,b}\nn\\[.5cm]
    &\quad +\fft18F_{\alpha\beta}^{(+)\,a}F_{\beta\gamma}^{(+)\,b}F_{\gamma\delta}^{(+)\, a} F_{\delta\alpha}^{(+)\,b} - \dfrac{1}{16} F_{\alpha\beta}^{(+)\,a} F_{\gamma\delta}^{(+)\,a} F_{\alpha\beta}^{(+)\,b} F_{\gamma\delta}^{(+)\,b}\nn\\
    &\quad-\fft14F_{\alpha\beta}^{(+)\,a} F_{\beta\gamma}^{(+)\,a}F_{\gamma\delta}^{(+)\,b}F_{\delta\alpha}^{(+)\,b}+2P_\alpha^{(+-)\,ab}P_\gamma^{(+-)\,ab}F_{\alpha\beta}^{(+)\,c}F_{\beta\gamma}^{(+)\,c}\nn\\
    &\quad-\fft32P_\alpha^{(+-)\,ba}P_\alpha^{(+-)\,ca}F_{\mu\nu}^{(+)\,b}F_{\mu\nu}^{(+)\,c}-4P_\gamma^{(+-)\,ba}P_\alpha^{(+-)\,ca}F_{\alpha\beta}^{(+)\,b}F_{\beta\gamma}^{(+)\,c}\nn\\
    &\quad-4P_\alpha^{(+-)\,ab}P_\beta^{(+-)\,ab}P_\alpha^{(+-)\,cd}P_\beta^{(+-)\,cd}+8P_\alpha^{(+-)\,ab}P_\beta^{(+-)\,cb}P_\alpha^{(+-)\,cd}P_\beta^{(+-)\,ad}\nn\\
    &\quad-8P_\alpha^{(+-)\,ab}P_\alpha^{(+-)\,cb}P_\beta^{(+-)\,cd}P_\beta^{(+-)\,ad}\nn\\[.5cm]
    &\quad+\fft12R_{\alpha\beta\gamma\delta}(\omega_+)F_{\alpha\beta}^{(+)\,a}F_{\gamma\delta}^{(+)\,a}-\dfrac{1}{8} F_{\alpha\beta}^{(+)\,a} F_{\beta\gamma}^{(+)\,b} F_{\gamma\delta}^{(-)\,a} F_{\delta\alpha}^{(-)\,b} \nn\\
    &\quad+ \dfrac{1}{16} F_{\alpha\beta}^{(+)\,a}F_{\gamma\delta}^{(+)\,a} F_{\alpha\beta}^{(-)\,b}F_{\gamma\delta}^{(-)\,b} -\fft12F_{\alpha\beta}^{(+)\,a}F_{\beta\gamma}^{(+)\,a}F_{\gamma\delta}^{(-)\,b}F_{\delta\alpha}^{(-)\,b} \nn \\
    &\quad-\fft14F_{\mu\nu}^{(+)\,a}F_{\mu\nu}^{(-)\,b}F_{\lambda\sigma}^{(+)\,a}F_{\lambda\sigma}^{(-)\,b} +2P_\alpha^{(+-)\,ab} P_\gamma^{(+-)\,ab} F_{\alpha\beta}^{(-)\,c} F_{\beta\gamma}^{(-)\,c}\nn \\ 
    &\quad-P_\alpha^{(+-)\,ab}P_\alpha^{(+-)\,ac}F_{\mu\nu}^{(-)\,b}F_{\mu\nu}^{(-)\,c} + 2P_\alpha^{(+-)\,ab}P_\gamma^{(+-)\,ac} F_{\alpha\beta}^{(-)\,b}F_{\beta\gamma}^{(-)\,c}\nn \\
    &\quad-3P_\gamma^{(+-)\,ab}P_\alpha^{(+-)\,ac}F_{\alpha\beta}^{(-)\,b}F_{\beta\gamma}^{(-)\,c}-\dfrac{1}{4}h_{\alpha\beta\gamma} h_{\alpha\delta\epsilon} F_{\beta\gamma}^{(+)\, a} F_{\delta\epsilon}^{(+)\, a}\nn\\
    &\quad+ \dfrac{1}{2} h_{\beta\alpha\gamma}h_{\beta\delta\epsilon} F^{(+)\, a}_{\alpha\delta} F^{(+)\, a}_{\gamma \epsilon} - h_{\alpha\beta\gamma} F^{(+)\,a}_{\alpha\delta} F^{(-)\, b}_{\gamma\delta} P_\beta^{(+-)\, ab}\nn\\
    &\quad+\dfrac{1}{2} h_{\alpha\beta\gamma} F^{(+)\,a}_{\alpha\delta} F^{(-)\, b}_{\beta\gamma} P_{\delta}^{(+-)\, ab} + \dfrac{1}{2} h_{\alpha\beta\gamma} F^{(-)\,b}_{\alpha\delta} F^{(+)\, a}_{\beta\gamma} P_{\delta}^{(+-)\, ab}.
\end{align}
%

\bibliographystyle{JHEP}
\bibliography{cite}

\end{document}